\documentclass{article} 
\usepackage{graphicx}
\usepackage{epstopdf, epsfig}
\usepackage[utf8]{inputenc}
\usepackage{natbib}
\usepackage{lineno,mathrsfs,gensymb}
\usepackage{amsmath,amssymb}
\usepackage{setspace,xcolor}
\usepackage{lscape,comment,multirow,appendix}
\usepackage{maths_commands}
\usepackage{color,cancel,fancyhdr}
\usepackage[margin=1in]{geometry}

\newcommand{\gpc}[1]{{\color{black}#1}}
\newcommand{\rev}[1]{{\color{black}#1}}
\newcommand{\revtwo}[1]{{\color{black}#1}}

\fancypagestyle{plain}{
    \fancyhf{}
    \rhead{Regimes of stratified turbulence at low Prandtl number}
    \lhead{Shah et al.}
    \rfoot{Page \thepage}

}
\title{Regimes of stratified turbulence at low Prandtl number}

\author{Kasturi Shah$^{1,2}$, Gregory P. Chini$^{3,4}$, Colm-cille P. Caulfield$^{1,5}$ and Pascale Garaud$^{6}$}

\date{\small
$^{1}$ Department of Applied Mathematics and Theoretical Physics, University of Cambridge,
Cambridge CB3 0WA, UK \\
$^{2}$ Department of Earth, Atmospheric and Planetary Sciences, Massachusetts Institute of Technology, Cambridge MA 02139, USA \\
$^{3}$ Program in Integrated Applied Mathematics, University of New Hampshire, Durham, NH 03824, USA. \\
$^{4}$ Department of Mechanical Engineering, University of New Hampshire, Durham, NH 03824, USA \\
$^{5}$ Institute for Energy and Environmental Flows, University of Cambridge, Cambridge CB3 0EZ, UK \\
$^{6}$ Department of Applied Mathematics, Baskin School of Engineering, University of California Santa
Cruz, Santa Cruz, CA 95064, USA
}

\begin{document}

\maketitle

\thispagestyle{empty} 

\begin{abstract}
Quantifying transport by strongly stratified turbulence in low Prandtl number ($Pr$) fluids is critically important for the development of better models for the structure and evolution of stellar \rev{and planetary} interiors. Motivated by recent numerical simulations showing strongly anisotropic flows suggestive of scale-separated dynamics, we perform a multiscale asymptotic analysis of the governing equations. We find that, in all cases, the resulting slow-fast system\rev{s} take a quasilinear form. Our analysis also reveals the existence of several distinct dynamical regimes depending on the emergent buoyancy Reynolds and P\'eclet numbers, $Re_b = \alpha^2 Re$ and $Pe_b = Pr Re_b$, respectively, where $\alpha$ is the aspect ratio of the large-scale turbulent flow structures, and $Re$ is the outer scale Reynolds number. Scaling relationships relating the aspect ratio, the characteristic vertical velocity, and the strength of the stratification (measured by the Froude number $Fr$)  naturally emerge from the analysis. When $Pe_b \ll \alpha$, the dynamics at all scales is dominated by buoyancy diffusion, and  our results recover the scaling laws empirically obtained from direct numerical simulations by \citet{cope2020}. For $Pe_b \ge O(1)$, diffusion is negligible (or at least subdominant) at all scales and our results are consistent with those of \citet{chini2022} for strongly stratified geophysical turbulence at $Pr = O(1)$. Finally, we have identified a new regime for $\alpha \ll Pe_b \ll 1$, in which slow, large scales are diffusive while fast, small scales are not. We conclude by presenting a map of parameter space that clearly indicates the transitions between isotropic turbulence, non-diffusive stratified turbulence, diffusive stratified turbulence and viscously-dominated flows\rev{, and by proposing parameterisations of the buoyancy flux, mixing efficiency and turbulent diffusion coefficient for each regime.}
\end{abstract}

\section{Introduction}
\label{sec:intro}

Quantifying vertical transport (of heat, chemical tracers and momentum) in the stably stratified regions of 
\rev{stars and gas giant planets}
is paramount to a better understanding of their structure and \rev{evolution}, 
as the 
transport rates required to match observations 
are largely inconsistent with purely diffusive processes. The same 
\rev{challenge arises in the modeling of}  stably stratified regions of the Earth's atmosphere and oceans. The solution in both cases is to include turbulent transport in the models, but this in turn requires reliable parametrizations of small-scale fluxes as functions of the local properties of the fluid and of its large-scale flow \citep{Munk1966,Pinsonneault1997,Iveyal2008,Greggetal2018,Aertsal2019, Caulfield2021}. 


Relatively small-scale turbulent vertical mixing in a stably stratified region typically incurs an energetic cost to \rev{raise} dense fluid parcels irreversibly and correspondingly lower  buoyant parcels \rev{and thus} can only be sustained on long time scales by \rev{mechanisms that} tap into some larger-scale energy reservoir. \rev{Fluid instabilities comprise one broad class of such mechanisms, with d}ifferent fluid instabilities access\rev{ing distinct} sources of energy: for instance, baroclinic and double-diffusive instabilities draw from the available potential energy of the fluid, magnetic instabilities rely on the magnetic energy of a finite-amplitude large-scale field, and shear instabilities \rev{of a large-scale mean flow, or of the motions associated with internal waves,} can tap into the kinetic energy reservoir\rev{s} of \rev{those flows}. The nature of the turbulence driven by each of these instabilities differs substantially, so that it ``still remains extremely difficult to say anything generic about [stratified] mixing" \citep{Caulfield2021}.


In this work, we \rev{restrict our attention to} modeling \rev{stratified} turbulence in stellar \rev{and planetary} interiors driven exclusively by shear. 
\rev{Astrophysical} fluids generally have a low Prandtl number $Pr$, where
 \begin{equation}
Pr=  \frac{\nu^*}{\kappa^*}, 
\end{equation}
$\nu^*$ is the \rev{(dimensional)} kinematic viscosity of the fluid, and $\kappa^*$ is the \rev{corresponding (dimensional)} buoyancy diffusivity (e.g. thermal or compositional diffusivity, depending on the main source of stratification); see figure 7 of \citet{Garaudal2015}.
\rev{Here, and throughout this work,} 
dimensional quantities are starred, while non-dimensional quantities are not.
We ignore rotation, magnetic fields, \rev{externally-driven large-scale internal waves}, and the presence of multiple components contributing to the flow buoyancy. Although these effects undoubtedly are important in most 
situations, the dynamics of purely shear-driven stratified turbulence in low $Pr$ fluids is still far from well understood, justifying the \rev{apparently} narrow scope of this study.

\rev{Under these simplifying assumptions, the principal remaining source of turbulent vertical transport in our model is a vertical shear, expected to lead to `overturning' horizontally-aligned vortices.} 
Balancing the potential energy cost and the kinetic energy gain of adiabatic turbulent eddies in a vertically sheared flow, \citet{Richardson1920} concluded that turbulence can \rev{only} be sustained provided 
\begin{equation}
J = \frac{N^{*2}}{S^{*2}}  \le O(1),  
\label{eq:Richcrit_mean}
\end{equation}
where $N^*$ is the typical value of the buoyancy frequency of the stratification, and $S^*$ is the typical vertical shearing rate of the flow.  Condition (\ref{eq:Richcrit_mean}) is known today as the Richardson criterion (where the nondimensional parameter $J$  is commonly referred to as a gradient Richardson number, which can vary with the vertical position $z^*$). 
For linear normal mode disturbances in an inviscid, non-diffusive parallel shear flow, \citet{Miles1961} and \citet{Howard1961} formalized this argument to establish rigorously the necessary condition for linear instability to be that $J(z^*) < 1/4$ somewhere within the flow. 

The Richardson criterion can be relaxed when the flow disturbances are not purely adiabatic, because the energy cost of vertical motions is reduced if the advected fluid parcels radiatively \citep{Townsend1958,Dudis1973} or diffusively \citep{zahn1974,Jones1977,Lignieresal1999} adjust their buoyancy to that of the background stratification on a time scale comparable with, or shorter than, their turnover time scale. 
In the diffusive case, this adjustment is sufficiently fast when the eddy P\'eclet number 
\begin{equation}
Pe_{\ell} \equiv \frac{u_{\ell}^* \ell^*}{\kappa^*}  \le O(1) ,
\label{eq:Pedef}
\end{equation}
 where $u_{\ell}^*$ is the characteristic velocity of an eddy of size $\ell^*$. \citet{zahn1974} heuristically argued that an appropriate criterion for the instability of diffusive eddies of size $\ell^*$ in a vertical shear flow is  
\begin{equation}
J Pe_{\ell} \le O(1) ,
\label{eq:loPecrit}
\end{equation}
provided (\ref{eq:Pedef}) holds. 
According to his criterion, it ought to be possible to maintain turbulence in strongly stratified shear flows ($J \gg 1$) provided eddies are small enough to ensure that $Pe_{\ell} \le O(J^{-1})$. 

\rev{The criterion (\ref{eq:loPecrit}) for these so-called `diffusive' shear instabilities should not be used in geophysical fluids where $Pr \ge O(1)$ because} the condition $Pe_{\ell} \le O(1)$ equivalently implies that  $Re_{\ell} \equiv u_{\ell}^* \ell^*/\nu^* \le O(1)$, where $Re_{\ell}$ is the corresponding eddy Reynolds number. As shear instabilities must have a relatively large Reynolds number to develop (otherwise viscous energy losses are too great), the requirement that $Re_{\ell} \gg 1$ is incompatible with $Pe_{\ell} = Pr Re_{\ell} \le O(1)$ when $Pr \ge O(1)$. 
The situation is quite different in stellar \rev{and planetary} interiors, as first noted by \citet{SpiegelZahn1970} and \citet{zahn1974}, because of their intrinsically small Prandtl number. With $Pr \ll 1$, and regardless of the size of the outer scale Reynolds and P\'eclet numbers 
\begin{equation}
Re \equiv \frac{U^*L^* }{\nu^*} \mbox{  and    }  Pe \equiv \frac{ U^*L^*}{\kappa^*},
\end{equation}
(where $U^*$ and $L^*$ are the system-scale characteristic flow velocity and length scale, respectively), there is always an intermediate range of scales where turbulent eddies satisfy $Pe_{\ell} \ll 1 \ll Re_{\ell}$, namely, where diffusion dominates their dynamics while viscous forces remain negligible.
 This property naturally allows for the maintenance of diffusive stratified turbulence provided the shear is large enough for (\ref{eq:loPecrit}) to hold \citep{zahn1974}.


\rev{An important question is the source of the vertical shear.  Naively, one might expect it to arise from the spatial gradients of a mean flow on large vertical scales, which are commonly observed in geophysical and astrophysical fluids. However, these large-scale `mean' gradients generally have corresponding  Richardson numbers that are much larger than unity, especially in stellar or planetary interiors \citep{garaud2021}. This naive expectation would also fail to explain observations of stratified turbulence in laboratory experiments \citep{Ruddickal1989,Parkal1994,HolfordLinden1999,Oglethorpeal2013}
and numerical experiments \citep{jacobitz2000,basak2006, brethouwer2007,maffioli2016,Lucasal2017,zhou2017,ZhouDiamessis2019,cope2020,garaud2020,Yi2023}, where the mean flow has no vertical shear. 

Instead, it is the \textit{emergent, instantaneous, local} vertical shear that is responsible for much of the vertical mixing in the flow \revtwo{\citep{riley2003}}. The emergent shear intermittently appears and disappears as different horizontal layers of the fluid move past each other, driven by primarily horizontal and highly anisotropic large-scale flows. Crucially, it has been shown that these vertically decoupled flow structures, which are sometimes called `pancake eddies', emerge {\it even when the horizontal flow is forced in a vertically-invariant manner}, as in the above-mentioned studies. Furthermore,} \gpc{in the stably stratified regions of the ocean, there is increasing evidence that \rev{the emergent shear is} often marginally stable \rev{to vertical shear instabilities}, with the minimum value of $J(z^*)\simeq 1/4$ \citep{Smyth2013}, suggestive of `self-organized criticality' \citep{Salehipour2018,Smyth2019,Mashayek2022}.} \rev{Similar results have recently been reported by \citet{Garaudal2024a} in numerical simulations of strongly stratified turbulence at $Pr = 0.1$. As demonstrated by \citet{chini2022} at $Pr = O(1)$, and in this paper at low $Pr$, this self-organized criticality seems to be an inevitable outcome of the highly anisotropic nature of the layerwise large-scale flows in strongly stratified fluids.} 


\rev{In the spirit of these previous studies,} 
\rev{we further narrow our focus to studying} vertical mixing \rev{in primarily horizontal flows driven} by 
\rev{a vertically-invariant horizontal force}, first at $Pr \ge O(1)$, where most of the research on this topic has been focused, and then at $Pr \ll 1$. \rev{First, however, we review the existing literature in some detail, in order to place our work in context.}

\subsection{Stratified turbulence \rev{forced} by horizontal flows at $Pr \ge O(1)$}

\rev{In studies} of stratified turbulence \rev{forced primarily by} horizontal flows\rev{, the effects of stratification generally are quantified} 
using the outer-scale Froude number, defined as 
\begin{equation}
Fr = \frac{U^*}{N^*L^*}.   
\end{equation}
In this expression $U^*$ and $L^*$ are now specifically assumed to be the characteristic \rev{outer scales for the} {\it horizontal} velocity and \rev{\it horizontal} length scale \rev{and over that outer scale there is also a well-defined (implicitly constant) `background' buoyancy frequency $N^*$}. 
\citet{Ruddickal1989}, \citet{Parkal1994} and \citet{HolfordLinden1999} demonstrated experimentally that regular horizontal motions of vertical rods in a salt-stratified fluid (having $Pr = O(1000)$) with an initially constant stratification, characterized by a buoyancy frequency $N^*$, can cause substantial vertical mixing even when \rev{the flow is `strongly' stratified, in the sense that} $Fr \ll 1$. Furthermore, they often observed the formation of steps in the density profile, with a characteristic vertical height $H^* \propto U^* / N^*$. \citet{Oglethorpeal2013} similarly found substantial mixing and the \rev{spontaneous} formation \rev{and long-time survival of a `staircase' of relatively well-mixed} layers \rev{with depth again proportional to $U^* / N^*$, separated by significantly thinner interfaces of locally-enhanced density gradient} in stratified Taylor-Couette flow experiments. 
 
Direct numerical simulations (DNS) are arguably more practical for investigation of $Pr = \mathit{O}(1)$ fluids (such as thermally-stratified air and water\rev{, due to the inherent difficulties associated with conductive heat losses at the boundaries}), and over the last 20 years significant progress in quantifying stratified turbulence in this parameter regime has been enabled by advances in supercomputing. In particular, access to the full three-dimensional structure of the velocity and  buoyancy fields has provided new insights into the nature of turbulence at very large Reynolds numbers and very strong stratification, 
a regime characteristic of
stratified turbulence in the ocean and atmosphere. 

Pioneering work by \citet{brethouwer2007}, for instance, has demonstrated that the eddy field is highly anisotropic in this regime, with a vertical eddy scale once again proportional to $U^*/N^*$
\rev{. This emergent `layer' length scale thus seems to be} a universal property of stratified turbulence at both moderate and high $Pr$ at sufficiently high Reynolds number, as theoretically predicted and experimentally demonstrated by  \cite{BillantChomaz2000,billant2001} \citep[see also][for a review]{Caulfield2021}.

Using the insight gained from their DNS data, \citet{brethouwer2007} proposed an anisotropic rescaling of the governing equations, introducing the small vertical scale $H^* = Fr L^*  =U^*/N^*$ and taking the asymptotic limit $Fr\to 0$. Crucially, the horizontal scales are assumed to remain $O(L^*)$ in their work. Inspection of the rescaled equations reveals the fundamental role of the so-called buoyancy Reynolds number 
\begin{equation}
Re_b \equiv \rev{Fr^2 Re = }   \left(\frac{H^*}{L^*}\right)^2 Re,
\end{equation}
 which needs to be substantially greater than one for viscous effects to be negligible \citep[see][and \S~\ref{sec:anisotropy} for further details]{brethouwer2007,BarthelloTobias2013}. In that case, the characteristic  vertical velocities and buoyancy fluctuations are predicted to scale as $W^* \propto Fr U^*$ and $B^* \propto Fr L^* N^{*2}=H^* N^{*2}$, respectively. \rev{This flow regime where
 $Re \gg 1$ and $Fr  \ll 1$ 
 such that $Re_b \equiv Fr^2 Re \gg 1$ was referred to as  the `strongly stratified regime, sometimes abbreviated to stratified turbulence' by \cite{brethouwer2007}. The  nomenclature `strongly stratified turbulence' (SST)  has become quite common in the fluid dynamics literature,
 although `stratified turbulence' is often (particularly in the geosciences literature) used more broadly to refer to any disordered motions in a stratified environment, as opposed to this particular asymptotic regime.
 
 Interestingly and importantly, \cite{brethouwer2007} presented an alternative expression for $Re_b$ in terms of the kinetic energy dissipation rate  $\varepsilon^*$, by assuming an effective definition for the horizontal outer scale $L^*$ such that
 \begin{equation}
     \varepsilon^* \equiv \frac{U^{*3}}{L^*} \, \rightarrow \,
 Re_b \equiv \frac{\varepsilon^*}{\nu^* N^{*2}} .
 \label{eq:rebeps}
 \end{equation}
This expression reveals $Re_b$ to be
an `intensity' or `activity' parameter for the turbulence \citep{gibson80,gargett84}.
\revtwo{Considered in this fashion, a particular attraction of the  SST nomenclature becomes apparent, as it succinctly describes the \emph{necessary} properties of the bulk flow, i.e. that the flow is `strongly stratified' in the sense that $Fr \ll 1$ when using global or bulk measures for the outer length and velocity scales and the buoyancy frequency and yet the flow is also `turbulent' in the sense that the disorder is sufficiently `intense' or `active' to be considered turbulent.}

However, \revtwo{there is emerging evidence that this} classical inertial scaling  may only be applicable when the effects of buoyancy are sufficiently weak \citep{Mashayek2022}, also perhaps calling into question whether it is appropriate to think of  such flows as  `strongly stratified'.} 
 
Fundamental to the arguments presented in \cite{brethouwer2007} is the \rev{notion} that there is a single (large, outer) horizontal scale of significance \rev{(indeed they referred to the assumption leading to (\ref{eq:rebeps}) as `the perhaps most important assumption' for the forward cascade to small scales essential to their theory)}, with the implicit consequence that spatio-temporal variability on smaller horizontal scales does not exert a controlling influence on the dynamics. 
\rev{The} supposition \rev{of a single dominant horizontal scale, however,} is not entirely self-consistent. 
\rev{Indeed,} \cite{brethouwer2007} demonstrated that an inevitable consequence of their scaling is that $J \simeq O(1)$ at least somewhere in the flow, thus implying 
\rev{that they are locally weakly stratified,
and potentially unstable to fast, small-scale isotropic perturbations. The coexistence of
strongly stratified ($J\gg 1$) quiescent regions and weakly stratified ($J \sim O(1)$) turbulent
regions in $Fr \ll 1$ flows has led \citet{falder2016} to propose that this strongly stratified turbulent regime rather 
be referred to as the ‘layered anisotropic stratified turbulence’ (LAST) regime.
This nomenclature has the twin advantage that it is specifically descriptive, as the flow
velocities are inherently anisotropic and layered, and also avoids ambiguity concerning the actual sense in
which the flow is ‘strongly’ stratified, in that a small \revtwo{bulk or global} value of $Fr$ may well mask
local regions in space and/or time where the stratification is sufficiently ‘weak’ to allow  such unstable, fast small-scale isotropic perturbations to develop \citep[see also][for further discussion]{Portwood2016}.
}
\revtwo{That said, 
we acknowledge that
`strongly stratified turbulence' is more concise terminology and does not refer to flow properties which are emergent, such as the emergent anisotropy and inevitable spatio-temporal intermittency. Nevertheless, DNS results in \citet{Garaudal2024b} clearly do show two such emergent distinct flow behaviours as stratification increases (see their figure 3), i.e. turbulent patches dominated by small-scale isotropic motions and layer-like, quiescent flow outside these regions. It is this regime characterised by the coexistence of large and small scales that we refer to as the LAST regime.} 

\rev{Later, \citet{riley2012} employed local but anisotropic energy cascade arguments to deduce that $W^*$ actually peaks at \emph{small} scales and is proportional to $Fr^{1/2} U^*$ rather than $Fr U^*$. DNS of strongly stratified turbulence by \citet{brethouwer2007}, \citet{augier2015} and \citet{maffioli2016} do, indeed, exhibit spatio-temporally intermittent motions on small scales. Furthermore, the $W^* \propto Fr^{1/2} U^*$ scaling law was tentatively verified by \citet{maffioli2016}.}

More recently, \citet{chini2022} 
\rev{leveraged the scale separation observed in these DNS to propose} a new multiscale asymptotic model of stratified turbulence at $Pr = O(1)$, using the concept of marginal stability to constrain the representative minimum values of the gradient Richardson number to be $O(1)$, thus determining key properties of `fast' motions associated with the presumed breakdown of these local shear instabilities. Importantly, this model  recovers the usual (and empirically observed) scaling law $H^* \propto Fr L^*$ for the vertical eddy scale, but predicts a characteristic vertical velocity scale $W^* \propto Fr^{1/2} U^*$ and a characteristic buoyancy scale $B^* \propto 
\rev{H^* N^{*2}}$. 
\rev{This} scaling for $W^*$ 
differ\rev{s} from the corresponding predictions of \citet{brethouwer2007} as a direct consequence of relaxing
the assumption of a single horizontal length scale to allow for the idealized modelling of certain important aspects of the spatiotemporally intermittent shear instabilities.
\rev{Moreover, the analysis by \citet{chini2022} also differs from that of \citet{riley2012} in three key aspects. First, in \citet{chini2022}, the scaling of $W^*$ arises from a self-consistent asymptotic theory rather than by invoking heuristic Kolmogorov turbulence arguments. Second, the scaling behaviour is associated with a spectrally \textit{nonlocal} energy cascade from large to small horizontal scales, whereas \citet{riley2012} rely on spectrally local cascade arguments that presume $\alpha$  
\revtwo{increases from its minimum value at the large-scale, $\alpha \sim Fr$, to unity at the Ozmidov scale, $\alpha \sim 1$, where isotropy is recovered.}
Third, the vertical velocity scaling in \citet{chini2022} is realised at horizontal scales of $\mathit{O}(H^*) = \mathit{O}(Fr L^*)$, which is asymptotically larger than the Ozmidov scale $l_O^* \propto Fr^{3/2} L^*$ at which \citet{riley2012} conclude it arises. Data from recent DNS by \citet{Garaudal2024b} tentatively support the argument of \citet{chini2022} that the vertical kinetic energy spectrum of stratified turbulence peaks at a scale proportional to $Fr$.}

\subsection{Stratified turbulence \rev{forced} by horizontal flows at $Pr \ll 1$}

By contrast with geophysical applications, the study of stratified turbulence driven by \rev{primarily} horizontal flows in stellar \rev{and planetary} interiors is still in its infancy.
Most of the early work on the topic is summarized in the seminal paper of \citet{zahn1992}, who, \rev{had used arguments similar to those of \citet{riley2012} to} estimate the vertical 
turbulent diffusivity $D^*$ of momentum, or of a passive scalar, in terms of the \rev{kinetic} energy dissipation rate $\varepsilon^*$ and of the local fluid properties $\kappa^*$ and $N^*$: 
 \begin{equation}
D^* \propto \left( \frac{\varepsilon^* \kappa^*}{N^{*2}} \right)^{1/2} .
\label{eq:ZahnD}
\end{equation}
To arrive at this conclusion, he made two assumptions: 
\rev{(1) the energy spectrum of the anisotropic horizontal scales of the flow follows a classical inertial scaling, analogously to (\ref{eq:rebeps}) such that  $\varepsilon^* \propto u^{*3}_{\ell}/\ell^*$; and (2) the vertical diffusivity is primarily due to the largest eddies whose emergent vertical shear $S^*$ is unstable according to (\ref{eq:loPecrit}), where $S^* = u^*_{\ell}/\ell^*$.} 
\citet{lignieres2021} recently revisited Zahn's argument, interpreting the relevant scale $\ell^*$ as a `diffusive' or `modified' Ozmidov scale and explicitly writing it as
 \begin{equation}
 \ell^*_{OM} = \left( \frac{\kappa^* \varepsilon^{*1/3}}{N^{*2}}\right)^{3/8}. 
 \end{equation}
Using critical balance theory, \citet{Skoutnev2023} then argued that it is possible to relate $\ell^*_{OM}$ and $u^*(\ell^*_{OM})$ explicitly to properties of the larger-scale flow ($U^*$, $W^*$, $H^*$ and $L^*$), arriving at the conclusion that 
\begin{subequations}
\begin{equation}
H^* \propto \left( \frac{U^*\kappa^*}{N^{*2} L^{*3} }\right)^{1/4} L^* = \left( \frac{Fr^2}{Pe}\right)^{1/4} L^*, \mbox{   and  } W^* \propto \left( \frac{Fr^2}{Pe}\right)^{1/4} U^*, 
\label{eq:lzskoutnev}
\end{equation}
so 
\begin{equation}
D^* \propto H^* W^* \propto \left( \frac{Fr^2}{Pe}\right)^{1/2} U^* L^* =\left( \frac{U^{*3}\kappa^*}{L^*N^{*2}}\right)^{1/2},
\label{eq:Dturbintro}
\end{equation}
\end{subequations}
which is consistent with Zahn's estimate (\ref{eq:ZahnD}) provided one further assumes that $\varepsilon^* \propto  U^{*3}/L^*$. 

As turbulence in stars cannot be directly observed, and experimentation with $Pr \ll 1$ fluids is particularly difficult, numerical simulations are best suited to provide insight into the low $Pr$ regime. \citet{cope2020} and \citet{garaud2020} recently presented a series of DNS of stratified turbulence driven by horizontal shear at low $Pr$, focusing on flows where the outer scale P\'eclet number $Pe$ is low and high, respectively. They found, as in the $Pr = O(1)$ simulations of \citet{brethouwer2007}, that the turbulence becomes highly anisotropic as stratification increases ($Fr$ decreases), and identified three distinct anisotropic regimes: (1) a fully turbulent regime, (2) a regime where the turbulence is spatially and temporally intermittent (as viscosity begins to affect the flow), and (3) a fully viscous regime (where viscosity completely controls the flow dynamics). In the fully turbulent and intermittent regimes, the horizontal motions contain well-separated large and small scales. 

When $Pe \ll 1$, diffusion is important at all scales, regardless of the stratification. \citet{cope2020} found empirically \rev{from their DNS} that the vertical size and characteristic velocity of eddies in the fully turbulent regime 
\rev{have the following apparent dependence on non-dimensional parameters} 
\begin{equation}
 H^* \propto  \left( \frac{Fr^2}{Pe}\right)^{1/3} L^*,\qquad  W^* \propto \left( \frac{Fr^2}{Pe}\right)^{1/6} U^*,
\label{eq:lzcope}
\end{equation}
\rev{for which a rigorous self-consistent theory 
has yet to be developed.} 
\rev{In particular, \eqref{eq:lzcope}} notably differ\rev{s} from the predictions of \citet{Skoutnev2023}. This discrepancy between the DNS results of \citet{cope2020} and Skoutnev's theory needs to be explained and is one of the principal motivations for this paper. Interestingly, the prediction for the turbulent diffusivity $D^* \propto H^*W^*$ yields the {\it same} expression (\ref{eq:Dturbintro}) as in \citet{zahn1992}, \citet{lignieres2021} and \citet{Skoutnev2023}, despite $H^*$ and $W^*$ satisfying different scaling laws. 

It is important to note, however, that \citet{zahn1992} did not distinguish between viscous and non-viscous regimes or diffusive and non-diffusive regimes, assuming instead that the turbulence can always develop on scales where diffusion is important but viscous effects are negligible. Yet, the results of \citet{cope2020} and \citet{garaud2020} demonstrate that viscosity always eventually begins to affect the turbulence as the stratification increases. 
By analogy with \citet{brethouwer2007} and \citet{chini2022}, one may expect the buoyancy Reynolds number $Re_b$ to be the relevant bifurcation parameter describing the impact of viscosity, a result that will be formally established in this paper. \citet{garaud2020} also demonstrated numerically the existence of a regime of stratified turbulence at $Pe \gg 1$ and $Pr \ll 1$ where diffusion is negligible. This is not surprising in hindsight, but was not anticipated by \citet{zahn1992}. She tentatively proposed that her data is consistent with $H^* \propto Fr^{2/3} L^*$, and $W^* \propto Fr^{2/3} U^*$ as long as viscosity remains negligible, which is \rev{perhaps} surprising given that one might naively have expected to recover the geophysical regime scalings in that case. She acknowledged, however, that her simulations may not have been performed at a sufficiently high Reynolds number to be in a meaningful asymptotic regime yet. 
\rev{Recently, \citet{Garaudal2024b} revisited this dataset and demonstrated that the inferred $W^* \propto Fr^{2/3} U^*$ scaling is an artefact of the coexistence of turbulent patches where $W^* \propto Fr^{1/2}U^*$, and more quiescent regions where $W^* \propto Fr U^*$.}  

The results presented above illustrate the complexity of stratified turbulence at low $Pr$, and motivate the need for additional work 
to determine
how many regimes exist, how salient flow properties scale with input parameters in each regime, and finally, where the regime boundaries lie in parameter space. In this paper, we therefore approach the problem systematically by adapting the multiscale asymptotic methodology developed by \citet{chini2022} for the $Pr  = O(1)$ regime to the low $Pr$ regime. 
We begin in \S~\ref{sec:anisotropy} by laying out the model equations and boundary conditions and performing the anisotropic scaling analysis \rev{in} \citet{brethouwer2007} \rev{on the $Pr \ll 1$ case}, explicitly comparing \rev{it to} the $Pr = O(1)$ case. This analysis reveals the crucial role of the buoyancy P\'eclet number (thus named by analogy with $Re_b$), \emph{viz.} $Pe_b = Pr Re_b$. 
We then perform a slow-fast decomposition of the governing equations in \S~\ref{sec:multiscale}. At high $Pe_b$, we recover the results of \citet{chini2022} and highlight the reason why this multiscale approach leads to a different vertical velocity scaling from the \rev{single-scale} one derived by \cite{brethouwer2007}. We then present new results at low $Pe_b$. Our findings are discussed in \S~\ref{sec:disc}, where we use this analysis to partition the parameter space into various regimes of stratified turbulence at low $Pr$ and explicitly provide scaling relationships for the vertical length scale and velocity in each case. Implications and potential applications of our results are presented in \S~\ref{sec:conclusions}. 

\section{Governing equations and anisotropic scalings }
\label{sec:anisotropy}

Consider a three-dimensional, non-rotating, incompressible flow expressed in a Cartesian coordinate system where the vertical coordinate $z^*$ (with unit vector $\hat{\boldsymbol{e}}_z$) is anti-aligned with gravity $({\bf g}^* = -g^* \hat{\boldsymbol{e}}_z)$. The fluid is stably stratified, with mean density $\rho_0^*$ and constant background buoyancy frequency $N^*$.
Buoyancy perturbations away from this mean state are incorporated in accordance with the Boussinesq approximation \citep{spiegel1960}. 
A \rev{vertically-invariant divergence-free horizontal} body force \rev{$\boldsymbol{F}^*_h$, 
which only varies slowly with time and space,} 
is applied to drive a mean horizontally sheared flow \rev{in a domain of size $(L^*_x,L^*_y,L^*_z)$}. 
The dimensional governing equations are
\begingroup
 \allowdisplaybreaks
 \begin{subequations}\label{eqn:spiegel_veronis}
\begin{alignat}{2}
     \pd{\boldsymbol{u^*}}{t^*} + \boldsymbol{u^*} \cdot  \boldsymbol{\nabla}^*  \boldsymbol{u^*}  &= - \frac{1}{\rho_0^*} \nabla^* p^* + b^* \hat{\boldsymbol{e}}_z + \nu^* \nabla^{*2} \boldsymbol{u^*}  
     + \frac{\rev{\boldsymbol{ F}^*_h}}{\rho_0^*}, \\
     \boldsymbol{\nabla}^* \cdot \boldsymbol{u^*}  &= 0, \\
     \pd{b^*}{t^*} + \boldsymbol{u^*} \cdot \boldsymbol{\nabla}^* b^* + N^{*2} w^* &= \kappa^* \nabla^{*2} b^* ,
 \end{alignat}
 \end{subequations}
 \endgroup
where \rev{$t^*$ is time,} $\boldsymbol{\nabla^*} = (\partial/\partial x^*,\partial/\partial y^*,\partial/\partial z^*)$ \rev{where $x^*$ and $y^*$ are horizontal spatial coordinates}, $\boldsymbol{u^*} = (u^*,v^*,w^*)$  denotes the velocity field, $p^*$ the pressure, and $b^*$ the buoyancy perturbation with respect to the  background stratification. In accord with the Boussinesq approximation, the fluid has a constant kinematic viscosity $\nu^*$ and constant diffusivity $\kappa^*$. 

\rev{ 
\subsection{Anisotropic single-scale equations for $Pr=\mathit{O}(1)$} \label{sec:highPrani} } 


In the limit of strong stratification, vertical displacements are energetically costly, and hence,
as discussed in \S\ref{sec:intro}, fluid motions become strongly anisotropic (with $u^*, v^* \gg w^*$). 
 Therefore, it seems natural to  non-dimensionalize the governing equations \eqref{eqn:spiegel_veronis}
anisotropically. \citet {billant2001} and \citet{brethouwer2007} have argued that the (dimensional) horizontal and vertical length scales of turbulent eddies ought to be $L^*$ and $H^*$, respectively, with an aspect ratio
\begin{equation}
    \alpha \equiv \frac{H^*}{L^*} \ll 1,
\end{equation} 
where the dependence of $\alpha$ on the stratification (and other governing parameters) is determined from the following asymptotic analysis.
We then introduce a new vertical coordinate $\zeta^*$ such that
 \begin{equation}
     z^* =  \alpha \zeta^*.
 \end{equation}
Consequently, if the horizontal velocity scale is $U^*$, then the vertical velocity scale must be $\alpha U^*$ to respect the divergence free condition without overly restricting the allowable types of flows.
 Time should be scaled by the turnover time of the horizontal eddies $L^*/U^*$ and pressure by $\rho_0^* U^{*2}$. To ensure that the nonlinear terms balance the forcing, $U^*= (F^*_0 L^*/\rho_0^*)^{1/2}$, where $F_0^*$ is the characteristic forcing amplitude. Finally, the buoyancy scale is chosen to be $H^* N^{*2}$ to enable the vertical advection of the background stratification to enter the buoyancy equation at leading order and to balance the horizontal advection of the buoyancy fluctuations.
 
Denoting the horizontal components of the velocity as $\pp{\boldsymbol{u}} = (u,v)$, and similarly the horizontal gradient as $\pp{\boldsymbol{\nabla}} = (\partial/\partial x,\partial/\partial y)$, the dimensionless system is given by
\begin{subequations}\label{eqn:full_dimensionless_eqn}
\begingroup
\allowdisplaybreaks
\begin{alignat}{2}
& \pd{\pp{\boldsymbol{u}}}{t} + \left( \pp{\boldsymbol{u}} \cdot \nabla_\perp \right) \pp{\boldsymbol{u}} +   w \pd{\pp{\boldsymbol{u}}}{\zeta} = - \pp{\nabla} p + \frac{1}{Re \alpha^2} \left( \alpha^2 \pp{\nabla}^2 \pp{\boldsymbol{u}} + \pd{^2 \pp{\boldsymbol{u}}}{\zeta^2} \right) + \rev{\boldsymbol{F}_h}, 
\label{eq:anisox1}\\
    & \pd{w}{t} + \left( \pp{\boldsymbol{u}} \cdot \nabla_\perp \right) w +   w \pd{w}{\zeta} = -\frac{1}{\alpha^2} \pd{p}{\zeta} + \frac{b}{Fr^2}  + \frac{1}{Re\alpha^2} \left( \alpha^2 \pp{\nabla}^2 w + \pd{^2 w}{\zeta^2} \right), \\
  & \pp{\nabla} \cdot \pp{\boldsymbol{u}} +   \pd{w}{\zeta} = 0, \\
&\pd{b}{t} + \left( \pp{\boldsymbol{u}} \cdot \nabla_\perp \right) b + w \pd{b}{\zeta} + w = \frac{1}{Pe \alpha^2} \left( \alpha^2 \pp{\nabla}^2b + \pd{^2 b}{\zeta^2} \right) \label{eq:anisob1},
\end{alignat}
\endgroup
\end{subequations}
where all variables are now non-dimensional and 
\begin{equation}
    Re = \frac{U^*L^*}{\nu^*}, \quad 
    Fr  = \frac{U^*}{N^*L^*}, \quad 
    Pe = \frac{U^*L^*}{\kappa^*} = Pr Re
\end{equation}
are the usual 
Reynolds, Froude, 
and P\'eclet numbers based on the characteristic horizontal scales of the flow. 

When the stratification is strong, $Fr \ll 1$, and the buoyancy term in the vertical component of the momentum equation is unbalanced unless it is compensated by the vertical pressure gradient. 
In other words, the flow anisotropy implies that hydrostatic balance must be satisfied at lowest order, which then requires 
\begin{equation}
\label{eq:highPralpha}
    \alpha  = Fr, 
\end{equation} 
implying that the characteristic vertical velocity $W^* = FrU^*$ while the characteristic vertical scale of the flow $H^* = U^*/N^*$. As discussed in \S~\ref{sec:intro}, this scaling for $H^*$ is well-established, and has been observed in several laboratory and numerical experiments  
\citep{HolfordLinden1999,brethouwer2007,Oglethorpeal2013}. 

Keeping only the lowest-order terms in an asymptotic expansion of \eqref{eqn:full_dimensionless_eqn} in $\alpha = Fr \ll 1$ yields
\begin{subequations}\label{eqn:highPrani}
\begin{alignat}{2}
    &\pd{\pp{\boldsymbol{u}}}{t} + \left( \pp{\boldsymbol{u}} \cdot \nabla_\perp \right) \pp{\boldsymbol{u}} +   w \pd{\pp{\boldsymbol{u}}}{\zeta} = - \pp{\nabla} p + \frac{1}{Re_b} \pd{^2 \pp{\boldsymbol{u}}}{\zeta^2} + \rev{\boldsymbol{F}_h}
    , \\
    &\pd{p}{\zeta} =  b, \quad \quad 
\pp{\nabla} \cdot \pp{\boldsymbol{u}} +    \pd{w}{\zeta} = 0, \tag{\ref{eqn:highPrani}\textit{b},\textit{c}} \\
    &\pd{b}{t} + \left( \pp{\boldsymbol{u}} \cdot \nabla_\perp \right) b +  w \pd{b}{\zeta} +  w = \frac{1}{Pe_b }  \pd{^2 b}{\zeta^2}, \tag{\ref{eqn:highPrani}\textit{d}}
\end{alignat}
\end{subequations}
where 
\begin{equation}
\label{eq:Reb}
    Re_b = \alpha^2 Re
\end{equation}
is the usually-defined buoyancy Reynolds number and 
\begin{equation}
\label{eq:Peb}
    Pe_b = Pr Re_b =  \alpha^2 Pe
\end{equation} 
is the corresponding buoyancy P\'eclet number. Note that when $Pr = O(1)$, the condition $Re_b \ge O(1)$, which is necessary for viscous effects to be small or negligible, implies that $Pe_b = Pr Re_b \ge O(1)$. As such, the effects of buoyancy diffusion are also small or negligible. 
The set of equations \eqref{eqn:highPrani}, which will be referred to as the ``anisotropic \rev{single-scale} high-$Pe_b$" equations hereafter \rev{(``single-scale high-$Pe_b$" equations for short)}, recovers the results of \citet{billant2001} in the inviscid and non-diffusive limit and of \citet{brethouwer2007} when viscous and diffusive effects are incorporated.

\subsection{Anisotropic \rev{single-scale} equations for $Pr \ll 1 $}
\label{sec:lowPrani}


When the Prandtl number is asymptotically small, it is possible to have a regime where $Pe_b \ll 1 \le Re_b$. In this extreme limit, the diffusion term on the right-hand-side of \eqrefl{eqn:full_dimensionless_eqn}{d} becomes unbalanced, unless the buoyancy field itself is much smaller than anticipated by the scaling $H^*N^{*2}$ used in the previous section. 
The strongly diffusive limit has in fact already been studied by \citet{lignieres1999}, who  showed that the buoyancy equation asymptotically reduces to a balance between the vertical advection of the background stratification and the diffusion of the buoyancy fluctuations. In dimensional terms, we therefore expect $N^{*2} w^* \simeq \kappa^* \nabla^2 b^*$. 
With this in mind, we let $b = Pe_b \hat b$ in (\ref{eq:anisox1})--(\ref{eq:anisob1}), and anticipate that $\hat b =O(1)$ (this is equivalent to scaling the dimensional buoyancy by $\alpha^3 N^{*2} U^*L^{*2}/\kappa^*$ instead of $H^*N^{*2}$). The resulting dimensionless system becomes 
\begin{subequations}
\label{eqn:dimensionless_eqn}
\begingroup
\allowdisplaybreaks
\begin{alignat}{2}
    &\pd{\pp{\boldsymbol{u}}}{t} + \left( \pp{\boldsymbol{u}} \cdot \nabla_\perp \right) \pp{\boldsymbol{u}} +  w \pd{\pp{\boldsymbol{u}}}{\zeta} = - \pp{\nabla} p + \frac{1}{Re_b} \left( \alpha^2 \pp{\nabla}^2 \pp{\boldsymbol{u}} + \pd{^2 \pp{\boldsymbol{u}}}{\zeta^2} \right) 
    + \rev{\boldsymbol{F}_h}
    , \label{eq:anisox2} \\
    &\pd{w}{t} + \left( \pp{\boldsymbol{u}} \cdot \nabla_\perp \right) w + w \pd{w}{\zeta} = -\frac{1}{\alpha^2} \pd{p}{\zeta} + \frac{Pe_b}{Fr^2} \hat b + \frac{1}{Re_b} \left( \alpha^2 \pp{\nabla}^2 w + \pd{^2 w}{\zeta^2} \right), \\
    &\pp{\nabla} \cdot \pp{\boldsymbol{u}} +  \pd{w}{\zeta} = 0, \\
    &\pd{\hat b}{t} + \left( \pp{\boldsymbol{u}} \cdot \nabla_\perp \right) \hat b + w \pd{\hat b}{\zeta} + \frac{1}{Pe_b} w = \frac{1}{Pe_b} \left( \alpha^2 \pp{\nabla}^2 \hat b + \pd{^2 \hat b}{\zeta^2} \right) \label{eq:anisob2}.
\end{alignat}
\end{subequations}
\endgroup
For sufficiently strong stratification \rev{(i.e. $Fr^2 \ll Pe_b$)}, the vertical component of the momentum equation must again be in hydrostatic balance, which requires 
\begin{equation}
\label{eq:lowPralpha}
    \alpha^2 = \frac{Fr^2}{Pe_b} = \frac{Fr^2}{\alpha^2 Pe}  \rightarrow \alpha = Fr_M , \mbox{ 
   where } Fr_M = \left(\frac{Fr^2}{Pe}\right)^{1/4}
\end{equation}
is a modified Froude number \citep[see][]{lignieres2021,Skoutnev2023}. 
We therefore find that 
the characteristic vertical length scale should be $H^* = (Fr^2/Pe)^{1/4} L^*$ and the characteristic vertical velocity scale should be $W^* = (Fr^2/Pe)^{1/4} U^*$, recovering the results of \citet{Skoutnev2023} albeit using a different argument.  

In the limit $Fr_M \ll 1$ and $Pe_b \ll 1$, keeping only the lowest order terms in (\ref{eq:anisox2})--(\ref{eq:anisob2}) yields
\begin{subequations}
\label{eqn:LPNani}
\begin{alignat}{2}
    &\pd{\pp{\boldsymbol{u}}}{t} + \left( \pp{\boldsymbol{u}} \cdot \nabla_\perp \right) \pp{\boldsymbol{u}} +  w \pd{\pp{\boldsymbol{u}}}{\zeta} = - \pp{\nabla} p + \frac{1}{Re_b}  \pd{^2 \pp{\boldsymbol{u}}}{\zeta^2} + \rev{\boldsymbol{F}_h}
    , \\
    & \pd{p}{\zeta} =  \hat b , \qquad \pp{\nabla} \cdot \pp{\boldsymbol{u}} +  \pd{w}{\zeta} = 0, \qquad w =  \pd{^2 \hat b}{\zeta^2} \tag{\ref{eqn:LPNani}\textit{b},\textit{c},\textit{d}}.
\end{alignat}
\end{subequations}
These scaling laws and governing equations are the low $Pe_b$ analogs of equations \eqref{eqn:highPrani}. In what follows, we therefore refer to them as the ``anisotropic \rev{single-scale} low-$Pe_b$" equations \rev{(``single-scale low-$Pe_b$" equations for short)}.  

\subsection{\rev{Motivation} for \rev{considering} multiscale dynamics}\label{sec:multiscale_evidence} %

\rev{For strongly stratified flows,} the reduced, anisotropic \rev{single-scale} equations given by \eqref{eqn:highPrani} for $Pe_b \ge O(1)$ and \eqref{eqn:LPNani} for $Pe_b \ll 1$ assume, by construction, that horizontal scales are large, while the vertical scale is small \rev{(e.g. in the $Pe_b \ge O(1)$ case, for strongly stratified flows, $\alpha \sim Fr$ necessarily implies $\alpha \ll 1$)}. 
As such, these equations describe the dynamics of  weakly-coupled ``pancake" vortices or horizontal meanders of the mean flow. They cannot, however, capture the small-scale turbulence that is expected to develop from shear instabilities between these layerwise horizontal motions \citep{chini2022}. Yet, these instabilities are ubiquitous when $Re_b$ is sufficiently large, and have been observed in laboratory experiments \rev{\citep{augier2014}} and inferred from oceanographic \emph{in situ} measurements \rev{\citep{falder2016}}
as well as in direct numerical simulations at $Pr \gtrsim O(1)$ \rev{\citep{riley2003,waite2011,augier2012}}. 

\rev{Indeed,} recent DNS of stratified turbulence 
has confirmed the presence and importance of flows on small horizontal scales. \rev{Small-scale features are clearly visible in figure 16 of \citet{maffioli2016},  figure 6 of \citet{cope2020},  figure 1 of \citet{garaud2020}, and figure 3 of \citet{Garaudal2024b}.} 
\rev{Furthermore,}
\rev{\citet{cope2020}} 
empirically find that $H^* \propto (Fr^2/Pe)^{1/3} L^*$ instead of $H^* \propto (Fr^2/Pe)^{1/4} L^*$ and that $W^* \propto (Fr^2/Pe)^{1/6} U^*$ instead of $W^* \propto (Fr^2/Pe)^{1/4} U^*$. These results thus \rev{call for further work to clarify} \rev{the conditions (if any) where}
the 
\rev{single-scale low-$Pe_b$} equations \eqref{eqn:LPNani} \rev{are relevant} for modeling stratified turbulence, at least at large buoyancy Reynolds number.

As discussed in \S~\ref{sec:intro}, \citet{chini2022} have recently argued in the context of geophysical stratified turbulence that one {\it must} take into account the fast, small horizontal scales and study their (marginally-stable) interaction with the slow, larger-scale anisotropic flow to obtain a more complete and more accurate model of stratified turbulence. Accordingly, we now \rev{hypothesise that multiscale dynamics is the missing physics required to obtain a self-consistent model of low $Pr$ stratified turbulence and} propose to extend their work to the low $Pr$ limit. The next section first outlines the work of \citet{chini2022} for pedagogical clarity, then extends the analysis to the low $Pr$ regime.

\section{Multiscale models for stratified turbulence}\label{sec:multiscale}

We consider the same model set-up as introduced in \S~\ref{sec:anisotropy}. Here, however,  we make no assumption about the amplitude of the vertical motions when non-dimensionalising the governing equations, and instead allow the appropriate scaling to emerge naturally from the analysis. Accordingly, 
we non-dimensionalize the vertical velocity by $U^*$ and correspondingly choose the buoyancy scale to be $L^*N^{*2}$. Then, the dimensionless system is
\begin{subequations}\label{eqn:full_dimensionless_eqn_iso}
\begingroup
\allowdisplaybreaks
\begin{alignat}{2}
& \pd{\pp{\boldsymbol{u}}}{t} + \left( \pp{\boldsymbol{u}} \cdot \nabla_\perp \right) \pp{\boldsymbol{u}} + \frac{w}{\alpha}   \pd{\pp{\boldsymbol{u}}}{\zeta} = - \pp{\nabla} p + \frac{1}{Re_b} \left( \alpha^2 \pp{\nabla}^2 \pp{\boldsymbol{u}} + \pd{^2 \pp{\boldsymbol{u}}}{\zeta^2} \right) + \rev{\boldsymbol{F}_h},
\\
    & \pd{w}{t} + \left( \pp{\boldsymbol{u}} \cdot \nabla_\perp \right) w +  \frac{w}{\alpha}  \pd{w}{\zeta} = -\frac{1}{\alpha} \pd{p}{\zeta} + \frac{b}{Fr^2}  + \frac{1}{Re_b} \left( \alpha^2 \pp{\nabla}^2 w + \pd{^2 w}{\zeta^2} \right), \\
  & \pp{\nabla} \cdot \pp{\boldsymbol{u}} + \frac{1}{\alpha} \pd{w}{\zeta} = 0, \\
&\pd{b}{t} + \left( \pp{\boldsymbol{u}} \cdot \nabla_\perp \right) b + \frac{w}{\alpha}  \pd{b}{\zeta} + w = \frac{1}{Pe_b} \left( \alpha^2 \pp{\nabla}^2b + \pd{^2 b}{\zeta^2} \right).
\end{alignat}
\endgroup
\end{subequations}
These equations are the starting point for our analysis. We assume that $Re_b \ge O(1)$, and that $Fr$ is sufficiently small to ensure that $\alpha \ll 1$, but make no other {\it a priori} assumption on the size of $Pe_b$ at this stage. 

\subsection{Slow-fast decomposition}

We now perform a multiscale expansion of the system in the limit of small aspect ratio $\alpha$. Following \citet{chini2022}, we assume the existence of two distinct sets of horizontal length scales: the original large scales that are $O(1)$ in the chosen nondimensionalization, as well as much smaller horizontal scales that are $O(\alpha)$. With that choice, small-scale fluid motions are isotropic by construction. 
We further assume that the flow has two distinct time scales: a slow time scale associated with the turnover of the large horizontal eddies, as before, and a fast time scale inversely related to the \rev{emergent} vertical shearing rate of the large-scale mean flow, $U^*/H^*$, \rev{that develops from the vertically-decoupled horizontal flows}. In practice, we thus define the slow and
fast horizontal coordinates as $\boldsymbol{x}_{s} = \pp{\boldsymbol{x}}$ and $\boldsymbol{x}_{f} = \boldsymbol{x}_{s}/\alpha$, respectively (henceforth, the subscript $f$ denotes fast and $s$ denotes slow). 
Similarly, we split time into slow and fast variables, such that $t_f = t_s/\alpha$ where $t_s = t$. 
Consequently, the partial derivatives with respect to time and to the horizontal variables become 
\begin{equation}\label{eqn:two_scale_t_x}
    \pd{}{t} = \frac{1}{\alpha} \pd{}{t_f} + \pd{}{t_s} , \qquad \pp{\nabla} = \frac{1}{\alpha} \nabla_{f} + \nabla_{s}. 
\end{equation}
All dependent variables (collectively referred to as $q$) are now assumed to be functions of both fast and slow length and time scales: $q = q(\boldsymbol{x}_{f},\boldsymbol{x}_{s},\zeta,t_f,t_s;\alpha)$. 
 
Assuming the fast and slow scales are sufficiently separated, \citet{chini2022} then define a fast-averaging operator $\overline{( \cdot )}$, such that 
\begin{equation}\label{eqn:fast_avg}
    \overline{q}(\boldsymbol{x}_{s},\zeta,t_s;\alpha) = \frac{1}{l_x l_y T} \int_\rev{\cal T} \int_{\cal D} q (\boldsymbol{x}_{ f},\boldsymbol{x}_{s},\zeta,t_f,t_s;\alpha) \mathrm{d} \boldsymbol{x}_{f} \mathrm{d} t_f,
\end{equation}
where ${\cal D}$ is a horizontal domain \rev{of size $(l_x \times l_y)$ centred on $\boldsymbol {x}_s$} 
\rev{where $\alpha \ll l_x,l_y \ll 1$}, and \rev{$\cal T$ is a temporal domain of size $\alpha \ll T \ll 1$ centred on $t_s$.} 
With this definition, $\overline{q}$ depends on slow variables only. Each quantity  
$q$ can then be split into a slowly-varying field $\bar q$ and a rapidly fluctuating component $q' = q-\overline{q}$, which implies that 
 the fast-average of the fluctuation field must vanish, i.e., $\overline{q'} = 0$. Note that $q'$ itself can still depend on the slow length and time scales.

We first substitute the expressions \eqref{eqn:two_scale_t_x} for $\pp{\nabla}$ and $\partial/\partial t$ into \eqref{eqn:full_dimensionless_eqn_iso}, and split each field $q$ into $\bar q + q'$. We then take the fast average of each of the four governing equations to obtain the evolution equations for the mean flow, then subtract the mean from the total to obtain the evolution equations for the fluctuations. 

Starting with the continuity equation, we have
\begin{equation}
    \frac{1}{\alpha} \nabla_{f} \cdot \boldsymbol{u'}_{\perp} +  \nabla_{s} \cdot \boldsymbol{\bar u}_{\perp} + \nabla_{s} \cdot \boldsymbol{u'}_{\perp} + \frac{1}{\alpha} \pd{\bar w}{\zeta} +  \frac{1}{\alpha} \pd{w'}{\zeta} = 0, 
\end{equation}
whose fast-average reveals that 
\begin{subequations}
\label{eqn:slowfast_cont}
\begin{equation}
\label{eqn:slowfast_cont_mean}
 \nabla_{s} \cdot \boldsymbol{\bar u}_{\perp}  + \frac{1}{\alpha} \pd{\bar w}{\zeta}  = 0
\end{equation}
for the mean flow and 
\begin{equation}
\label{eqn:slowfast_cont_fluct}
    \frac{1}{\alpha} \nabla_{f} \cdot \boldsymbol{u'}_{\perp}  + \nabla_{s} \cdot \boldsymbol{u'}_{\perp} +  \frac{1}{\alpha} \pd{w'}{\zeta} = 0
\end{equation}
\end{subequations}
for the perturbations.

A similar procedure for the horizontal momentum equation yields 
\begin{subequations}
\label{eqn:slowfast_u}
\begin{alignat}{2}
\label{eqn:slowfast_u_mean}
\pd{\overline{\boldsymbol{u}}_\perp}{t_s} &+ \overline{\boldsymbol{u}}_\perp \cdot \nabla_s \overline{\boldsymbol{u}}_\perp + \frac{\overline{w}}{\alpha}  \pd{\overline{\boldsymbol{u}}_\perp}{\zeta} + \frac{1}{\alpha} \left( \overline{ \boldsymbol{u}_\perp' \cdot \nabla_f \boldsymbol{u}_\perp' + w' \pd{\boldsymbol{u}_\perp'}{\zeta} } \right) + \overline{ \boldsymbol{u}_\perp' \cdot \nabla_s \boldsymbol{u}_\perp' } \nonumber \\
    &= - \nabla_s \overline{p} + \frac{1}{Re_b} \left( \pd{^2 \overline{\boldsymbol{u}}_\perp}{\zeta^2} + \alpha^2 \nabla_s^2 \overline{\boldsymbol{u}}_\perp \right) + \rev{ \bar{\boldsymbol{F}}_h},
\end{alignat}
for the mean flow (where 
\rev{the magnitude of the horizontal force is} $O(1)$ by construction), and 
\begin{alignat}{2}
\label{eqn:slowfast_u_fluct}
    \frac{1}{\alpha} \pd{\boldsymbol{u}_\perp'}{t_f}  &+\pd{\boldsymbol{u}_\perp'}{t_s} + \frac{1}{\alpha} \overline{\boldsymbol{u}}_\perp \cdot \nabla_f \boldsymbol{u}_\perp' + \overline{\boldsymbol{u}}_\perp \cdot \nabla_s \boldsymbol{u}_\perp' + \boldsymbol{u}_\perp' \cdot \nabla_s \overline{\boldsymbol{u}}_\perp + \frac{1}{\alpha} \boldsymbol{u}_\perp' \cdot \nabla_f \boldsymbol{u}_\perp' + \boldsymbol{u}_\perp' \cdot \nabla_s \boldsymbol{u}_\perp' \nonumber \\
    &+ \frac{1}{\alpha} \left( \overline{w} \pd{\boldsymbol{u}'_\perp}{\zeta} + w' \pd{\overline{\boldsymbol{u}}_\perp}{\zeta} + w' \pd{\boldsymbol{u}'_\perp}{\zeta} \right) = - \frac{1}{\alpha} \nabla_f p' \rev{- \nabla_s p'} + \frac{1}{Re_b} \left( \nabla^2_f \boldsymbol{u}'_\perp  + \pd{^2 \boldsymbol{u}'_\perp }{\zeta^2} \right) \nonumber\\
    &{\qquad\rev{+ \frac{\alpha^2}{Re_b} \nabla^2_s \boldsymbol{u}_\perp' + \frac{2 \alpha}{Re_b} \nabla_s \cdot \nabla_f \boldsymbol{u}_\perp' } + \frac{1}{\alpha} \left( \overline{ \boldsymbol{u}_\perp' \cdot \nabla_f \boldsymbol{u}_\perp' + w' \pd{\boldsymbol{u}_\perp'}{\zeta} } \right) + \overline{ \boldsymbol{u}_\perp' \cdot \nabla_s \boldsymbol{u}_\perp' }},
\end{alignat}
\end{subequations}
for the fluctuations. \rev{Note that $\bar {\boldsymbol F}_h =  {\boldsymbol F}_h$ because the forcing is slowly varying in time and space by assumption.} 

The mean buoyancy equation is
\begin{subequations}
\label{eqn:slowfast_b}
\begin{equation}\label{eqn:slowfast_b_mean}
    \pd{\overline{b}}{t_s} + \overline{\boldsymbol{u}}_\perp \cdot \nabla_s \overline{b} + \frac{\overline{w}}{\alpha}  \pd{\overline{b}}{\zeta} + \frac{1}{\alpha} \left( \overline{\boldsymbol{u}_\perp' \cdot \nabla_f b' + w' \pd{b'}{\zeta} } \right) + \overline{ \boldsymbol{u}_\perp' \cdot \nabla_s b' } + \overline{w} = \frac{1}{Pe_b} \left( \pd{^2 \overline{b}}{\zeta^2} + \alpha^2 \nabla_{s}^2 \overline{b} \right),
\end{equation}
 while the corresponding fluctuation equation becomes: 
\begin{alignat}{2}\label{eqn:slowfast_b_fluct}
    \frac{1}{\alpha} \pd{b'}{t_f} + \pd{b'}{t_s} & + \frac{1}{\alpha} \overline{\boldsymbol{u}}_\perp \cdot \nabla_f b' + \boldsymbol{u}_\perp' \cdot \nabla_s \overline{b} + \frac{1}{\alpha} \boldsymbol{u}_\perp' \cdot \nabla_f b' + \boldsymbol{u}_\perp' \cdot \nabla_s b' \nonumber \\
    &+ \frac{1}{\alpha} \left( w' \pd{\overline{b}}{\zeta} + \overline{w} \pd{b'}{\zeta} + w' \pd{b'}{\zeta} \right) + w' = \frac{1}{Pe_b}  \left( \nabla^2_f b'  + \pd{^2 b' }{\zeta^2} \right)\nonumber\\
    &{\rev{+ \frac{\alpha^2}{Pe_b} \nabla^2_s b' + \frac{2 \alpha}{Pe_b} \nabla_s \cdot \nabla_f b'} + \frac{1}{\alpha} \left( \overline{ \boldsymbol{u}_\perp' \cdot \nabla_f b' + w' \pd{b'}{\zeta} } \right) + \overline{ \boldsymbol{u}_\perp' \cdot \nabla_s b' }}.
\end{alignat}
\end{subequations}

Finally, the mean vertical momentum equation is 
\begin{subequations}
\label{eqn:slowfast_w}
\begin{alignat}{2}
\label{eqn:slowfast_w_mean}
    & \pd{\overline{w}}{t_s} + \overline{\boldsymbol{u}}_\perp \cdot \nabla_s \overline{w} + \frac{\overline{w} }{\alpha} \pd{\overline{w}}{\zeta} + \frac{1}{\alpha} \left( \overline{\boldsymbol{u}_\perp' \cdot \nabla_f w' + w' \pd{w'}{\zeta} } \right) + \overline{ \boldsymbol{u}_\perp' \cdot \nabla_s w' }  \nonumber \\  &
    = - \frac{1}{\alpha} \pd{\bar p}{\zeta} + \frac{\bar b}{Fr^2} +  \frac{1}{Re_b} \left(\pd{^2 \overline{w}}{\zeta^2} + \alpha^2 \nabla_s^2 \overline{w} \right),
\end{alignat}
while the fluctuations satisfy
\begin{alignat}{2}\label{eqn:slowfast_w_fluct}
    \frac{1}{\alpha} \pd{w'}{t_f} & + \pd{w'}{t_s}  + \frac{1}{\alpha} \overline{\boldsymbol{u}}_\perp \cdot \nabla_f w' + \boldsymbol{u}_\perp' \cdot \nabla_s \overline{w} + \frac{1}{\alpha} \boldsymbol{u}_\perp' \cdot \nabla_f w' + \boldsymbol{u}_\perp' \cdot \nabla_s w' \nonumber \\
    &+ \frac{1}{\alpha} \left( w' \pd{\overline{w}}{\zeta} + \overline{w} \pd{w'}{\zeta} + w' \pd{w'}{\zeta} \right)  = -\frac{1}{\alpha} \pd{p'}{\zeta} + \frac{b'}{Fr^2} +  \frac{1}{Re_b} \left( \nabla_f^2 w' + \pd{^2 w'}{\zeta^2} \right)\nonumber\\
    &{ \rev{+ \frac{\alpha^2}{Re_b} \nabla^2_s w' + \frac{2 \alpha}{Re_b} \nabla_s \cdot \nabla_f w'} + \frac{1}{\alpha} \left( \overline{ \boldsymbol{u}_\perp' \cdot \nabla_f w' + w' \pd{w'}{\zeta} } \right) + \overline{ \boldsymbol{u}_\perp' \cdot \nabla_s w' }}.
\end{alignat}
\end{subequations}

We see, as noted by \citet{chini2022}, that the effective Reynolds and P\'eclet numbers of the fluctuation equations are $Re_b/\alpha$ and $Pe_b / \alpha$, respectively, which implies that the fluctuations are formally much {\it less} viscous and {\it less} diffusive than the mean. This perhaps counterintuitive conclusion is a direct consequence of the flow anisotropy, \rev{which enables the fluctuations to evolve on a faster time scale than the mean}.

\subsection{Multiscale model at  $Pe_b  \ge O(1)$}
\label{sec:highPrmultiscale}

We begin by summarizing the steps taken by 
\citet{chini2022} to derive a reduced multiscale model for stratified turbulence at $Re_b,Pe_b \ge O(1)$, as much of the analysis proves to be similar at low Prandtl number. As in that work, we posit the following asymptotic expansions: 
\begin{equation}
\label{eqn:asym_expansions_highPr}
    [b,p,\pp{\boldsymbol{u}},w] \sim [b_0, p_0, \boldsymbol{u}_{\perp 0}, w_0] + \alpha^{1/2}[b_1, p_1, \boldsymbol{u}_{\perp 1}, w_1] + \alpha [b_2, p_2, \boldsymbol{u}_{\perp 2}, w_2] + \dots
\end{equation}
The expansions start at $O(1)$ to  reflect the expectation that the dominant contributions to the pressure and the horizontal velocity arise on large horizontal scales. 
Although the expansions for $b$ and $w$ also start at $\mathit{O}(1)$, we show below that $w_0$ and $b_0$ both vanish when $\alpha \rightarrow 0$. 
Finally, the expansions proceed as asymptotic series in $\alpha^{1/2}$ following the results of \citet{chini2022}. They demonstrated that, because the small-scale fluctuations are isotropic, $\boldsymbol{u'}$  and $w'$ are necessarily of the same order. Inspection of the mean horizontal momentum equation then immediately reveals that both fields need to be $O(\alpha^{1/2})$ to ensure that the Reynolds stresses feed back on $\boldsymbol{u}_{\perp 0}$ at leading order (see below for further details.) 

With these choices, we substitute \eqref{eqn:asym_expansions_highPr} into the equations with multiscale derivatives presented above, analysing in turn the continuity equation \eqref{eqn:slowfast_cont}, the horizontal component of the momentum equation \eqref{eqn:slowfast_u}, the buoyancy equation \eqref{eqn:slowfast_b}, and finally, the vertical component of the momentum equation \eqref{eqn:slowfast_w}. At each step, we match terms at leading order to infer their sizes and respective evolution equations, and thus derive a reduced model for the flow.

Considering first the mean continuity equation \eqref{eqn:slowfast_cont_mean}, we see that $\partial_\zeta\bar{w}_0=\partial_\zeta\bar{w}_1 = 0$, implying $\bar w_0 = \bar w_1 = 0$ 
to suppress unphysical `elevator modes' 
from our model. Consequently, 
\begin{subequations}
\label{eqn:highPr_multiscalecont}
\begin{equation}
    \nabla_{s} \cdot \boldsymbol{\bar u}_{\perp 0}  + \pd{\bar w_2}{\zeta} = 0.
    \end{equation}
In the fluctuation continuity equation \eqref{eqn:slowfast_cont_fluct}, the second term is clearly much smaller than the first and can therefore be neglected from the leading-order set of dominant terms. Substituting the asymptotic series \eqref{eqn:asym_expansions_highPr} we then see that
    \begin{equation}
 \nabla_{f} \cdot \boldsymbol{u'}_{\perp i}  + \pd{w'_i}{\zeta} = 0,
    \end{equation}
    for $i = 0, 1$ (for larger values of $i$, the slow derivative of $\boldsymbol{u'}_{\perp i-2}$ should be taken into account).   
\end{subequations}

We now turn to the mean horizontal component of the momentum equation \eqref{eqn:slowfast_u_mean}. The Reynolds stresses must be $O(1)$ to feed back on the mean flow, so $\boldsymbol{u'}$ and $w'$ must both be $O(\alpha^{1/2})$.
Hence, $\boldsymbol{u}_{\perp 0}' = \boldsymbol{0}$ and $w'_{0} = 0$, yielding $\boldsymbol{u}_{\perp 0} = \overline{\boldsymbol{u}}_{\perp 0}$ and that $w_{0} = 0$, since $\bar{w}_0=0$, too. \citep[See][for a more detailed discussion of why $\boldsymbol{u}_{\perp 0}' = \boldsymbol{0}$.]{chini2022} In the horizontal momentum equation for the fluctuations, 
the fast dynamics take place at $O(\alpha^{-1/2}$) since $\boldsymbol{u'} = O(\alpha^{1/2})$. Ensuring that pressure is a leading-order effect implies that $p'_0 = 0$, so $p_0 = \bar p_0$, as expected. Many of the remaining terms are formally higher order, including all fluctuation-fluctuation interactions, which are $O(1)$ or smaller. 
Of the nonlinear terms, the only ones that contribute at leading order are quasilinear: $\alpha^{-1} \overline{\boldsymbol{u}}_\perp \cdot \nabla_f \boldsymbol{u}'_\perp$ and $\alpha^{-1} w' \partial_z \overline{\boldsymbol{u}}_\perp$. Therefore, after substituting \eqref{eqn:asym_expansions_highPr} into \eqref{eqn:slowfast_u} and retaining only the leading terms, we obtain
\begin{subequations}
\label{eqn:highPr_multiscaleu}
\begin{alignat}{2}
\pd{\boldsymbol{\overline{u}}_{\perp 0} }{t_s} +  \boldsymbol{\overline{u}}_{\perp 0} \cdot \nabla_s  \boldsymbol{\overline{u}}_{\perp 0} + \overline{w}_2 \pd{\boldsymbol{\overline{u}}_{\perp 0} }{\zeta} &= - \nabla_{s} \overline{p}_0 - \pd{}{\zeta} \left( \overline{w_{1}' \boldsymbol{u}_{\perp 1}'} \right) +  \frac{1}{Re_b} \pd{^2 \boldsymbol{\overline{u}}_{\perp 0}}{\zeta^2} + \rev{\bar{\boldsymbol{F}}_h}, \\
\pd{\boldsymbol{u}_{\perp 1}'}{t_f} +  \boldsymbol{\overline{u}}_{\perp 0} \cdot \nabla_{f}  \boldsymbol{u}_{\perp 1}' + w_{1}' \pd{\boldsymbol{\overline{u}}_{\perp 0}}{\zeta}  &= - \nabla_{f} p_1' + \frac{\alpha}{Re_b}  \left( \nabla^2_f \boldsymbol{u}'_{\perp 1}  + \pd{^2 \boldsymbol{u}'_{\perp 1} }{\zeta^2} \right),
\end{alignat}
\end{subequations}
where the formally higher-order Laplacian term has been retained to regularize the fluctuation equation.

Next, we examine the buoyancy equation, which reveals the sizes of $\bar b$ and $b'$. In the mean equation \eqref{eqn:slowfast_b_mean}, $\bar w = O(\alpha)$, and the buoyancy flux term is formally also of that order \citep[see][and also below]{chini2022}. Thus, $\bar b = O(\alpha)$ as well, as long as $Pe_b \ge O(1)$, which is implicit in the regime considered in this section, implying $\bar b_0 = \bar b_1 = 0$. The size of the buoyancy flux can be confirmed by inspection of the buoyancy fluctuation equation \eqref{eqn:slowfast_b_fluct}.
To be in a regime in which the stratification impacts the turbulent motions, the $O(b'/\alpha)$ fast dynamics must be of the same order as the advection of the background stratification, which is $O(w') = O(\alpha^{1/2})$. This ordering implies that $b' = O(\alpha^{3/2})$, so $b'_0 = b'_1 = b'_2 = 0$. Consequently, $b_0 = b_1 =0$, and $b_2 = \bar b_2$. Fluctuation-fluctuation interactions in this equation are again formally higher order, and the only remaining nonlinearities are quasilinear. At leading order, using \eqref{eqn:asym_expansions_highPr} in \eqref{eqn:slowfast_b} shows that the mean and fluctuation buoyancy equations are  \begin{subequations}\label{eqn:highPr_multiscaleb}
 \begin{alignat}{2}
     \pd{\overline{b}_2}{t_s} + \boldsymbol{\overline{u}}_{\perp 0} \cdot \nabla_s \overline{b}_2  + \overline{w}_2 \pd{\overline{b}_2}{\zeta} &  + \overline{w}_2 = - \pd{}{\zeta} \left( \overline{w_{1}' b_3'} \right) +  \frac{1}{Pe_b} \pd{^2\overline{b}_2}{\zeta^2},  \\ 
              \pd{b_{3}'}{t_f} +  \boldsymbol{\overline{u}}_{\perp 0} \cdot \nabla_{f}  b_{3}' +  w_{1}' \pd{\overline{b}_2}{\zeta} & + w_{1}' =   \frac{\alpha}{Pe_b} \left( \nabla_f^2 b_{3}' + \pd{^2 b'_3}{\zeta^2} \right),      
 \end{alignat}
\end{subequations}
where as before the diffusion term in the fluctuation equation has been retained for regularization.
 
Finally, we identify the leading order terms in the vertical component of the momentum equation. In the mean equation \eqref{eqn:slowfast_w_mean}, a hydrostatic leading-order balance requires $\alpha = Fr$, as in the anisotropic \rev{single-scale high-$Pe_b$} model described in \S~\ref{sec:highPrani}. This specification, when applied to the fluctuation equation \eqref{eqn:slowfast_w_fluct}, is quite satisfactory as it implies that the fluctuating buoyancy force $b'/Fr^2$ arises at leading order. \revtwo{One implication of the $\alpha = Fr$ scaling is that the largest horizontal scale to exhibit isotropic dynamics is predicted to be $\mathit{O}(U^*/N^*)$, which is much larger than the Ozmidov scale (since $\ell_O=\mathit{O}(Fr^{1/2}U^*/N^*)$). Physically, this prediction is consistent with the preferential excitation of stratified shear instabilities having horizontal length scales comparable to the layer thickness, as demonstrated theoretically in \citet{chini2022} and numerically in \citet{augier2015}. Nevertheless, the Ozmidov scale retains its significance as the largest horizontal scale to be (largely) unaffected by buoyancy. }
It can also easily be shown that, once again, fluctuation-fluctuation interactions are negligible at leading order, so the fast dynamics are quasilinear. Substituting \eqref{eqn:asym_expansions_highPr} into \eqref{eqn:slowfast_w} and using all of the information available, we obtain the leading order mean and fluctuating components of the vertical momentum equation,
\begin{subequations}\label{eqn:highPr_multiscalew}
 \begin{alignat}{2}
     \pd{\overline{p}_0}{\zeta} &= \bar b_2, \\
     \pd{w_{1}'}{t_f} +  \boldsymbol{\overline{u}}_{\perp 0} \cdot \nabla_{f}  w_{1}' &= - \pd{p_1'}{\zeta} + b_{3}' + \frac{\alpha}{Re_b} \left(  \nabla_f ^2 w_{1}' + \pd{^2 w'_1}{\zeta^2} \right), 
     \end{alignat}
 \end{subequations}
where the formally higher-order viscous term has been retained to regularize the fluctuation equation.

 The system of equations formed by 
 \eqref{eqn:highPr_multiscalecont} (with $i=1$),  \eqref{eqn:highPr_multiscaleu}, \eqref{eqn:highPr_multiscaleb} and \eqref{eqn:highPr_multiscalew}, is equivalent to equations (2.28)--(2.35) in \citet{chini2022}, the only differences arising from a different choice of non-dimensionalization.
 \rev{Furthermore, we see that the mean flow equations (\ref{eqn:highPr_multiscaleu}a), (\ref{eqn:highPr_multiscaleb}a) and (\ref{eqn:highPr_multiscalew}a) recover the single-scale high-$Pe_b$ equations in (\ref{eqn:highPrani}) in the absence of fluctuations.}
The \rev{multiscale} equations are fully closed and therefore self-consistent. Crucially, the resulting system is quasilinear, and can be solved efficiently by appealing to (plausible and empirically supported) marginal stability arguments for the fluctuations as discussed by \citet{michel2019} and \citet{chini2022}.

These equations are valid whenever $\alpha \ll 1$ (equivalently, $Fr \ll 1$ since $\alpha = Fr$), and the buoyancy Reynolds and P\'eclet numbers are both $O(1)$ or larger, thus allowing for the growth and saturation of the fluctuations via interactions with the mean. When $Pr = \mathit{O}(1)$, $Re_b \ge \mathit{O}(1)$ necessarily implies $Pe_b \ge \mathit{O}(1)$, so the two conditions are equivalent. Since $Re_b = \alpha^2 Re$, this condition is equivalent to $Re,Pe \ge O(Fr^{-2})$. 

For small $Pr$, these reduced equations remain valid as long as $Re_b, Pe_b \geq O(1)$. However, for {\it sufficiently} small $Pr$ it is possible to have an intermediate regime where $Re_b \ge O(1)$ while $Pe_b = Pr Re_b \ll 1$, and that regime is {\it not} captured by the reduced equations derived here and in \citet{chini2022}.
Yet, this scenario is likely to be relevant in stellar interiors, where $Pr$ is asymptotically small \citep{Garaudal2015}. 
We now perform a similar analysis to
develop a multiscale model  that is valid for low-$Pe_b$ flows.

\subsection{Multiscale model for $Pe_b \ll 1$}

In the limit of small $Pe_b$, the arguments presented above continue to apply for the continuity equation and the horizontal component of the momentum equation, neither of which involve buoyancy terms. However, the procedure subsequently fails because diffusive effects dominate in the buoyancy equation and are unbalanced unless $b$ is much smaller than expected, mirroring the argument made in \S~\ref{sec:lowPrani}. To capture this correctly within the context of our proposed multiscale model, we must introduce a two-parameter expansion for small $\alpha$ {\it and} small $Pe_b$ in lieu of \eqref{eqn:asym_expansions_highPr}. 

As in \citet{chini2022}, we assume that the asymptotic expansion in $\alpha$ proceed\rev{s} in powers of $\alpha^{1/2}$ and show the self-consistency of this choice below. Inspired by \citet{lignieres1999}, we now also assume that each field can be expanded as a series in powers of $Pe_b$ as well. We therefore have 
\begin{alignat}{2}\label{eqn:two_param_exp}
     q = \, &  q_{00} + \alpha^{1/2} q_{01} + \alpha \, q_{02} + \dots  + 
    Pe_b \left( q_{10} + \alpha^{1/2} q_{11} + \alpha \, q_{12} + \dots \right) + O(Pe_b^2), 
\end{alignat}
for $q \in \{ \boldsymbol{u}_\perp, p, w, b\}$, and assume that $\boldsymbol{\bar u}_{\perp 00} = O(1)$ and $\bar p_{00} = O(1)$ to balance the forcing. We then proceed exactly as before, substituting \eqref{eqn:two_param_exp} in turn into the multiscale continuity equation \eqref{eqn:slowfast_cont}, the horizontal component of the momentum equation \eqref{eqn:slowfast_u}, the buoyancy equation \eqref{eqn:slowfast_b}, and the vertical component of the momentum equation \eqref{eqn:slowfast_w}, to extract the relevant reduced equations at leading order. 

Starting with the mean continuity equation \eqref{eqn:slowfast_cont_mean}, we find again that $\overline{w} = O(\alpha)$ and hence $\overline{w}_{00} = \overline{w}_{01} = 0$.
 Equation \eqref{eqn:slowfast_cont_fluct} further implies that $O(\boldsymbol{u}_\perp') = O(w')$, so 
 \begin{subequations}
\label{eqn:multiscale_cont}
\begin{alignat}{2}     
 & \nabla_{s} \cdot \boldsymbol{\overline{u}}_{\perp 00} + \pd{\overline{w}_{02}}{\zeta}  = 0 ,  \\     &\nabla_{f} \cdot \boldsymbol{u}_{\perp 01}' + \pd{w_{01}'}{\zeta} = 0. \end{alignat}
\end{subequations}

From the mean horizontal component of the momentum equation \eqref{eqn:slowfast_u_mean}, $\boldsymbol{u'}_{\perp}$ and $w'$ must both be $O(\alpha^{1/2})$ as before, hence $\boldsymbol{u}_{\perp 00}' = 0$ and $w'_{00} = 0$, so $\boldsymbol{u}_{\perp 00} = \overline{\boldsymbol{u}}_{\perp 00}$ and $w_{00} = 0$. 
The inferred sizes of the velocity fluctuations again shows that fluctuation-fluctuation interactions are formally higher-order, and the only remaining nonlinearities in the corresponding fluctuation equation \eqref{eqn:slowfast_u_fluct} are quasilinear. Finally, for the fluctuating horizontal pressure gradient to influence the fluctuations of horizontal velocity at leading order, $p' = O(\boldsymbol{u}_\perp') = O(\alpha^{1/2})$, hence $p_{00}' = 0$. Substituting \eqref{eqn:two_param_exp} into \eqref{eqn:slowfast_u_mean} and \eqref{eqn:slowfast_u_fluct}, using these deductions and retaining only the lowest-order terms, shows that
\begin{subequations}
\label{eqn:multiscale_u}
\begin{alignat}{2} \pd{\boldsymbol{\overline{u}}_{\perp 00} }{t_s} +  ( \boldsymbol{\overline{u}}_{\perp 00} \cdot \nabla_{s} ) \boldsymbol{\overline{u}}_{\perp 00} + \overline{w}_{02} \pd{\boldsymbol{\overline{u}}_{\perp 00} }{\zeta} & = - \nabla_{s} \overline{p}_{00} - \pd{}{\zeta} \left( \overline{w_{01}' \boldsymbol{u}_{\perp 01}'} \right) +  \frac{1}{Re_b} \pd{^2 \boldsymbol{\overline{u}}_{\perp 00}}{\zeta^2} + \rev{\bar{\boldsymbol{F}}_h},\\
\pd{\boldsymbol{u}_{\perp 01}'}{t_f} +  (\boldsymbol{\overline{u}}_{\perp 00} \cdot \nabla_{f} )  \boldsymbol{u}_{\perp 01}' + w_{01}' \pd{\boldsymbol{\overline{u}}_{\perp 00}}{\zeta} & = - \nabla_{f} p_{01}' + \frac{\alpha}{Re_b} \left(\nabla_f^2 \boldsymbol{u}_{\perp 01}' + \pd{^2 \boldsymbol{u}_{\perp 01}' }{\zeta^2} \right),  
\end{alignat}
\end{subequations}
where the viscous term in the fluctuation equation has been retained for regularization. 

Thus far, each step in the analysis has been identical to that taken in the previous section. Rapid diffusion, however, affects the size of the mean and fluctuating buoyancy  fields $\bar b$ and $b'$, and does so in different ways because the effective P\'eclet number of the fluctuations is larger than that of the mean flow. More specifically, having assumed in this section that $Pe_b \ll 1$, we see that two possibilities arise when $\alpha \ll 1$: either $\alpha \ll Pe_b \ll 1$, in which case diffusion is dominant in the mean buoyancy equation but negligible in the fluctuation buoyancy equation, or $Pe_b \ll \alpha$ in which case diffusion is dominant at all scales. In what follows, we investigate both cases in turn. 

Before doing so, however, we note that the mean buoyancy equation \eqref{eqn:slowfast_b_mean} in both cases is unbalanced unless  $\overline{b} = O(\alpha Pe_b)$ to match the $\bar w$ term (again mirroring the arguments given in \S~\ref{sec:lowPrani}). This implies $\overline{b}_{0i} = 0$, $\forall i$, and $\bar b_{10} = \bar b_{11} = 0$. At lowest order in $Pe_b$, the only surviving terms in \eqref{eqn:slowfast_b_mean} are therefore \begin{equation}\label{eqn:temp_b_mean}
    \frac{1}{\alpha} \left( \overline{\boldsymbol{u}_\perp' \cdot \nabla_f b' + w' \pd{b'}{\zeta} } \right) + \overline{w} = \frac{1}{Pe_b} \pd{^2 \overline{b}}{\zeta^2},
\end{equation}
where the size of $b'$ is yet to be determined and differs depending on the relative sizes of $Pe_b$ and $\alpha$. For this reason, we have retained the turbulent buoyancy flux for now.

\subsubsection{Case 1: $\alpha \ll Pe_b \ll 1$ (the intermediate regime)}
\label{sec:intermediate}

We first consider a scenario in which $\alpha \ll Pe_b \ll 1$ and henceforth refer to this part of parameter space as the ``intermediate regime". While diffusion dominates the mean buoyancy equation, the fact that $\alpha/Pe_b \ll 1$ implies that it only formally enters the 
buoyancy fluctuation equation \eqref{eqn:slowfast_b_fluct} at higher order. 
Because of this, the evolution of $b'$ is very similar to that obtained in \citet{chini2022}. As in \S~\ref{sec:highPrmultiscale}, we ensure that the background stratification influences the fast dynamics of $b'$ by requiring $b'/\alpha = O(w') = O(\alpha^{1/2})$, so $b' = O(\alpha^{3/2})$. This implies that $b_{00}', b_{01}', b_{02}' = 0$, but $b_{03}' \neq 0$. 

Substituting the ansatz \eqref{eqn:two_param_exp} into the mean and fluctuation buoyancy equations and using the information collected so far we therefore have 
\begin{subequations}
\label{eqn:intermediate_b}
    \begin{alignat}{2}
    & \pd{}{\zeta}
    \left( \overline{w'_{01} b'_{03}} \right) + \overline{w}_{02} =  \pd{^2 \overline{b}_{12}}{\zeta^2},\\
    &
    \pd{b'_{03}}{t_f} + \overline{\boldsymbol{u}}_{\perp 00} \cdot \nabla_f b'_{03}  +  w'_{01} &= \frac{\alpha}{Pe_b} \left( \nabla_f^2 b'_{03} + \pd{^2 b'_{03}}{\zeta^2}\right),
    \end{alignat}
\end{subequations}
where the diffusion term for the fluctuations can be retained to regularize the equation, but is formally higher order. 

We note that the reduced buoyancy fluctuation equation in this regime differs slightly from the one derived by \citet{chini2022} given in \eqrefl{eqn:highPr_multiscaleb}{b}, because it does not contain a term of the form $w'_{01} \partial \bar b/\partial \zeta$, which is formally higher-order when $Pe_b \ll 1$. Also, we see that the mean buoyancy equation in that regime differs from the asymptotic low P\'eclet number (LPN) equation of \citet{lignieres1999}, \rev{and from the one derived by \citet{Skoutnev2023}}, which \rev{do} not contain a turbulent flux term. This discrepancy arises because \rev{their} derivation assumes that all dynamics are diffusive, whereas in this intermediate regime the fluctuation dynamics are not and can therefore influence the mean buoyancy field at leading order. 

Finally, we examine the vertical component of the momentum equation. Based on past experience in the $Pe_b \ge O(1)$ case (see 
\S~\ref{sec:highPrmultiscale}), one would naively expect to recover hydrostatic equilibrium at leading order in the mean vertical momentum equation \eqref{eqn:slowfast_w_mean}. Because $\bar b = O(\alpha Pe_b)$, this would imply $\alpha = (Fr^2 /Pe_b)^{1/2}$ as in the 
\rev{single-scale low-$Pe_b$} equations of \S~\ref{sec:lowPrani}. However, that choice leads to an irreconcilable inconsistency in the fluctuation equation \eqref{eqn:slowfast_w_fluct}: the fluctuation pressure gradient term is $O(\alpha^{-1/2})$, as are the fast inertial dynamics of $w'$, but the fluctuation buoyancy term is $O(\alpha^{3/2}/Fr^2) = O(\alpha^{-1/2}/Pe_b)$, which is formally much larger than any other term and is therefore unbalanced. In \rev{other} words, we cannot reconcile hydrostatic equilibrium at leading order for the mean flow with a balanced equation for the fluctuation $w'$. 

The solution to this conundrum is to insist instead that the equation for $w'$ be balanced, in which case $O(\alpha^{-1/2} ) = O( \alpha^{3/2}Fr^{-2})$, thereby recovering the standard scaling relationship $\alpha = Fr$  \citep{billant2001,brethouwer2007,chini2022}. With this choice, the leading-order vertical pressure gradient in the mean equation is asymptotically small \citep[e.g. as for the wall-normal pressure gradient in laminar boundary-layer theory of][]{batchelor1967}.
This perhaps unexpected result is discussed in \S~\ref{sec:disc} \rev{and} below. Substituting \eqref{eqn:two_param_exp} into \eqref{eqn:slowfast_w}, we obtain the reduced mean and fluctuating vertical component of the momentum equation at leading order, 
\begin{subequations}
\label{eqn:intermediate_w}
\begingroup
\begin{alignat}{2}
& \pd{\bar p_{00}}{\zeta} = 0, \\
     \pd{w'_{01}}{t_f} + \boldsymbol{\overline{u}}_{\perp 00} \cdot \nabla_{f} w'_{01} &= -  \pd{p'_{01}}{\zeta} +  b'_{03} + \frac{\alpha}{Re_b} \left( \nabla_f^2  w'_{01} + \pd{^2 w'_{01}}{\zeta^2} \right), 
\end{alignat}
\endgroup
\end{subequations}
where the higher-order viscous term is added to regularize the fluctuation equation. 

The set of equations formed by \eqref{eqn:multiscale_cont}, \eqref{eqn:multiscale_u},  \eqref{eqn:intermediate_b}, and \eqref{eqn:intermediate_w} are the intermediate regime analogs of the reduced model given in \citet{chini2022}.
They are valid as long as $Re_b \ge O(1)$, and $\alpha \ll Pe_b \ll 1$. Given that $\alpha = Fr$ in this regime, this inequality constraint is equivalent to requiring that $Fr \ll 1$ (so $\alpha \ll 1$), $Re \ge Fr^{-2}$, and 
$Fr^{-1} \ll Pe \ll Fr^{-2}$. 
\rev{It is evident that, in the absence of fluctuations, these equations nearly recover the single-scale equations derived in \S~\ref{sec:lowPrani}, except that $\partial \bar p_{00}/\partial \zeta = 0$. 
Considering higher order terms in the vertical momentum equation, we successively find that $\partial_\zeta \bar{p}_{01} = 0$ as well; it is only at the next order that buoyancy influences the mean flow, i.e., $\partial_\zeta \bar{p}_{10} = \bar{b}_{12}$.
By replacing $\overline{p}_{00}$ in the vertical mean momentum equation with a composite pressure, $\bar{p}_c = \overline{p}_{00} + \alpha^{1/2} \bar{p}_{01} + Pe_b \, \bar{p}_{10}$, we
recover (in the absence of fluctuations) the single-scale low-$Pe_b$ equations and those in \citet{Skoutnev2023}. For consistency, the corresponding composite mean horizontal momentum equation should be derived accurate to $O(\alpha^{1/2},Pe_b)$ in terms of a composite mean horizontal velocity $\boldsymbol{\overline{u}}_{\perp c}$. 
This exercise, however, does not yield any structurally new terms that do \textit{not} involve fluctuations (but only higher-order corrections to existing mean terms in \eqrefl{eqn:multiscale_u}{a}). Consequently, in the singular limit of vanishing fluctuations, the multiscale model with a composite mean pressure reduces to the single-scale low-$Pe_b$ equations in \S~\ref{sec:lowPrani}. 
} 
The physical implication\rev{s} of equations \rev{\eqref{eqn:multiscale_cont}, \eqref{eqn:multiscale_u},  \eqref{eqn:intermediate_b}, and \eqref{eqn:intermediate_w}}, and their potential caveats, are discussed in \S~\ref{sec:disc}. 

\subsubsection{Case 2: $Pe_b \ll \alpha$ (the fully diffusive regime)}\label{sec:meanLowPeb_fluctLowPe_b}

We now consider the regime where $Pe_b \ll \alpha$, in which both mean and fluctuating buoyancy fields are dominated by diffusion. Accordingly, we refer to this part of parameter space as the fully diffusive regime. Inspection of \eqref{eqn:slowfast_b_fluct} shows that the diffusion term in the fluctuation equation is unbalanced unless $b' = O(Pe_b w')$. We previously found that the vertical velocity fluctuations are $O(\alpha^{1/2})$, which implies here that $b'=  O(\alpha^{1/2} Pe_b)$. We conclude that $b'_{0i} = 0$  $\forall i$, and that $b'_{10} = 0$ as well. 
Combined with the results obtained from analysis of the mean buoyancy equation in the diffusive limit, we conclude that $b_{10} = 0$ while $b_{11} = b'_{11}$.

Using this information and substituting \eqref{eqn:two_param_exp} into \eqref{eqn:slowfast_b},
we obtain at lowest order 
\begin{subequations}
\begingroup
\label{eqn:lowPeb_b}    
\begin{alignat}{2}
\label{eqn:lowPeb_b_mean}    
     & \overline{w}_{02} =  \pd{^2 \overline{b}_{12}}{\zeta^2}, \\
\label{eqn:lowPeb_b_fluct}
    & w'_{01} = \nabla^2_f b'_{11}  + \pd{^2 b'_{11} }{\zeta^2}, 
\end{alignat}
\endgroup
\end{subequations}
which is as expected from the LPN dynamics central to this regime \citep{lignieres1999}. The buoyancy equation is linear and does not contain any time dependence, instead instantaneously coupl\rev{es} the vertical velocity and buoyancy fields. 
The validity of Ligni\`eres' LPN equations was verified numerically by \citet{cope2020}, for instance.

As usual, the last step of the derivation involves analysis of the vertical component of the momentum equation. As in \S~\ref{sec:intermediate}, requiring hydrostatic equilibrium for the mean flow would imply $\alpha^2 = Fr^2 / Pe_b$ \citep{lignieres2021,Skoutnev2023}, but leads to an inconsistency in the fluctuation equation, where the buoyancy term  would be unbalanced. To see this, note that $b'/Fr^2 = O( \alpha^{1/2} Pe_b/Fr^2) = O( \alpha^{-3/2})$ with that choice for $\alpha$,
while the fluctuating pressure term and all other dominant terms in the equation are only  $O(\alpha^{-1/2})$. As in the previous section, the resolution to this inconsistency is to insist that the buoyancy term in the 
 fluctuation equation \rev{for $w'
 $} should be balanced instead. Here, this implies 
\begin{equation}
O\left( \frac{p'}{\alpha}\right)  =  O\left(\frac{b'}{Fr^2}\right)  \rightarrow  \alpha = \frac{Fr^2}{Pe_b},
\end{equation}
using the fact that $p' = O(\alpha^{1/2})$ and $b' = O(\alpha^{1/2} Pe_b)$. Recalling that $Pe_b = \alpha^2 Pe$, we then recover the crucial scaling relationship 
\begin{equation}
\label{eq:lowPebalpha}
\alpha = \left(\frac{Fr^2}{ Pe}\right)^{1/3} = Fr_M^{4/3},
\end{equation} 
which had originally been proposed by \citet{cope2020} based on their DNS data. Our multiscale analysis therefore provides a sound theoretical basis for their empirical results.

After substituting \eqref{eqn:two_param_exp} into \eqref{eqn:slowfast_w} and using the available information, we obtain
\begin{subequations}
    \label{eqn:lowPeb_w}
    \begin{alignat}{2}
    \label{eqn:lowPeb_w_mean}
    \pd{\bar p_{00}}{\zeta} = 0, \\ \label{eqn:lowPeb_w_fluct}
        \pd{w'_{01}}{t_f} + ( \boldsymbol{\overline{u}}_{\perp 00} \cdot \nabla_{f} ) w'_{01} & = -  \pd{p'_{01}}{\zeta} +  b'_{11} + \frac{\alpha}{Re_b} \left( \nabla_f^2  w'_{01} + \pd{^2 w'_{01}}{\zeta^2} \right), 
        \end{alignat}
\end{subequations}
which is very similar to the system obtained in the intermediate regime studied in \S~\ref{sec:intermediate}, except for the appearance of the buoyancy fluctuation term $b'_{11}$ instead of $b'_{03}$. As before, formally higher-order viscous terms are retained to regularize the fluctuation equation. 

The set of equations formed by \eqref{eqn:multiscale_cont}, \eqref{eqn:multiscale_u}, \eqref{eqn:lowPeb_b}, and \eqref{eqn:lowPeb_w} are the fully-diffusive regime analogs of the reduced model derived by \citet{chini2022}. 
They are valid as long as $Re_b \ge O(1)$, and $Pe_b \ll \alpha \ll 1$. Given that $\alpha = (Fr^2/Pe)^{1/3}$ in this regime, this parameter constraint is equivalent to requiring that $Fr^2 \ll Pe$ (to ensure $\alpha \ll 1$), $Pe \ll Fr^{-1}$ (to ensure $Pe_b \ll \alpha$), and $Pe \ge Pr^3 Fr^{-4}$ (to ensure that $Re_b \ge O(1)$). 
The three conditions demarcate a triangle in logarithmic parameter space in which the equations are valid -- see \S~\ref{sec:turbulent_regimes}. As an important self-consistency check, we see that the $Pe_b = \alpha$ transition between the fully diffusive and intermediate regimes is the same ($Pe = Fr^{-1}$) whether the transition is approached from the former or latter part of parameter space. 
\rev{Here, too, (weak) buoyancy effects can be included in the mean dynamics by considering higher-order pressure contributions to the vertical momentum equation. We successively find that $\partial_\zeta \bar{p}_{01} =0$ and $\partial_\zeta (\overline{w'_{01} w'_{01}}) = - \partial_\zeta \bar{p}_{02} + \bar{b}_{12}$. Hence, when fluctuations are absent from the system of multiscale equations given above, the resulting equations with an appropriately defined composite mean pressure $\overline{p}_c = \overline{p}_{00} + \alpha^{1/2} \bar{p}_{01} +  \alpha \, \overline{p}_{02}$ recover hydrostatic balance and the single-scale low-$Pe_b$ equations given in \S~2.2.} 
In the next section (\S~\ref{sec:disc}), we consider the physical implications of equations \rev{\eqref{eqn:multiscale_cont}, \eqref{eqn:multiscale_u}, \eqref{eqn:lowPeb_b}, and \eqref{eqn:lowPeb_w}} as well as potential caveats on their applicability.

\section{Discussion}
\label{sec:disc}

In this study, we have extended the results of \citet{chini2022} by performing a multiscale asymptotic analysis of stratified turbulence at low Prandtl number.
Our work demonstrates the existence of several different regimes depending on the strength of the stratification (quantified by the inverse Froude number) and the rate of buoyancy diffusion (quantified by the inverse P\'eclet number).
In each regime, the asymptotic analysis self-consistently yields  a slow-fast system of quasilinear equations describing the concurrent evolution of a highly anisotropic, slow, large-scale mean flow and isotropic, fast, small-scale fluctuations. \rev{The small scales are self-consistently excited by an instability of the emergent, local vertical shear. When they are not, the equations for the large-scale anisotropic flow recover exactly those obtained in \S~\ref{sec:anisotropy} (with a composite mean vertical momentum equation in the low $Pe_b$ case).} 

The large-scale anisotropy is characterized by the aspect ratio $\alpha$ (the ratio of the vertical to horizontal scales of the large-scale flow), whose functional dependence on $Fr$ and $Pe$ naturally emerges from the analysis. The various regime boundaries are  locations in parameter space where relevant Reynolds and P\'eclet numbers arising in the mean flow and fluctuation equations are $O(1)$, signifying transitions between viscous and (formally) inviscid dynamics, and/or diffusive and non-diffusive dynamics. We now summarize our findings in each regime and then discuss the model assumptions as well as the implications of our results for low Prandtl number fluids.

\subsection{Synopsis of multiscale equations and their validity}

The first regime is characterized by $Fr \ll 1$ and $Pe_b, Re_b \ge O(1)$, where $Re_b = \alpha^2 Re$ and $Pe_b = \alpha^2 Pe$. In this regime, we recover the reduced model of \citet{chini2022} and confirm that $\alpha = Fr$. Recalling that, at leading order, $w'_1 = w' / \alpha^{1/2}$, $b'_3 = b' / \alpha^{3/2}$, etc., that $\partial /\partial t_f = \alpha \partial /\partial t$ (and similarly \rev{$\nabla_f = \alpha \nabla_\perp$}), and finally that $\partial/\partial \zeta = \alpha \partial /\partial z$, we can rewrite \eqref{eqn:highPr_multiscalecont},  \eqref{eqn:highPr_multiscaleu}, \eqref{eqn:highPr_multiscaleb} and \eqref{eqn:highPr_multiscalew} as the following quasilinear system for mean and fluctuations, expressed in the original variables \rev{$(x,y,z,t)$}.
\\
 \noindent \textit{Mean flow equations:}
\begin{subequations}\label{eqn:highPr_multiscale}
 \begin{alignat}{2}
 \pd{\boldsymbol{\overline{u}}_{\perp} }{t} +  \boldsymbol{\overline{u}}_{\perp} \cdot \nabla\rev{_\perp} \boldsymbol{\overline{u}}_{\perp} \rev{+ \bar{w} \pd{\boldsymbol{\bar{u}}_\perp}{z} } &= - \nabla_{\perp} \overline{p} - \pd{}{z} \left( \overline{w' {\boldsymbol u'}_\perp} \right) +  \frac{1}{Re} \pd{^2 \boldsymbol{\overline{u}}_{\perp}}{z^2} + \rev{\bar{\boldsymbol{F}}_h},\\ 
     \pd{\overline{p}}{z} &= \frac{\bar b}{Fr^2},  \qquad \nabla \cdot \boldsymbol{\overline{u}} = 0 , \\ 
      \pd{\overline{b}}{t} +   \boldsymbol{\overline{u}}_{\perp} \cdot \nabla\rev{_{\perp}} \overline{b} \rev{\ + \bar{w} \pd{\bar{b}}{z} } + \overline{w} &  = -  \pd{}{z} \left( \overline{w' b'} \right) +  \frac{1}{Pe} \pd{^2\overline{b}}{z^2}.
 \end{alignat}
 
 \noindent \textit{Fluctuation equations:}
  \begingroup
 \allowdisplaybreaks
 \begin{alignat}{2}
     \pd{\boldsymbol{u}_{\perp}'}{t} +  \boldsymbol{\overline{u}}_{\perp} \cdot \nabla_{\perp}  \boldsymbol{u}_{\perp}' + w' \pd{\boldsymbol{\overline{u}}_{\perp}}{z} &= -  \nabla_{\perp} p' + \frac{1}{Re} \nabla^2 \boldsymbol{u}_{\perp}' , \\
     \pd{w'}{t} +  \boldsymbol{\overline{u}}_{\perp} \cdot \nabla_{\perp}  w' &= - \pd{p'}{z} + \frac{b'}{Fr^2} + \frac{1}{Re}  \nabla^2  w',  \\ 
	\nabla \cdot \boldsymbol{u}'&= 0 ,\\
         \pd{b'}{t} +  \boldsymbol{\overline{u}}_{\perp} \cdot \nabla_{\perp}  b' +  w' \pd{\overline{b}}{z} + w' &= \frac{1}{Pe} \nabla^2 b' .      
 \end{alignat}
 \endgroup
 \end{subequations}

This system, which is equivalent to equations (2.28)--(2.35) in \citet{chini2022}, is valid when $Fr \ll 1$, and  $Re,Pe \ge O(Fr^{-2})$, corresponding to $Re_b, Pe_b \geq O(1)$. 
\rev{Noting that the overbar denotes an intermediate averaging operation (see \eqref{eqn:fast_avg}), our reduced model has a generalised quasilinear (GQL) form \citep{MarstonChiniTobias2016,MarstonTobias2023}, albeit here derived in physical space (whereas the GQL reduction usually is performed in Fourier space). When slow horizontal variation is suppressed and the overbar denotes a strict horizontal average, the above reduced equations have the same form as an \textit{ad hoc} quasilinear (QL) reduction of the Boussinesq equations \citep[e.g.][] {Garaud2001,fitzgerald2018,fitzgerald2019}. In this work, we have demonstrated that this (G)QL form is in fact a natural outcome of the slow--fast asymptotic expansion and is therefore asymptotically exact in the given distinguished limit.}
\revtwo{As detailed in \citet{chini2022}, in 2D the fluctuation equations without diffusion can be reduced to the Taylor-Goldstein equation \citep[see for example][]{Craik1985} as the mean fields are independent of the fast variables $\boldsymbol{x}_{\perp f}$ and $t_f$ and hence are effectively frozen during the fast evolution of the fluctuations.
Thus, the fluctuation equations admit as solutions all of the `classical' stratified shear instabilities of parallel shear flows, including in particular Kelvin-Helmholtz instabilities. This mathematical reduction thus may explain the occurrence of Kelvin-Helmholtz billows within fully turbulent but strongly stratified flows. }

Dimensionally, this regime has a characteristic vertical scale  $H^* = \alpha L^* = U^*/N^*$. The characteristic vertical velocity is $W^* = \alpha^{1/2} U^* = (U^{*3}/N^*L^*)^{1/2}$, and $w$ is dominated by small-scale fluctuations. The characteristic buoyancy scale is $N^{*2} H^*$, and $b$ \rev{is} dominated by large scales. Note that the horizontal field \rev{$\boldsymbol{u}_\perp$} is predominantly large scale, by assumption, but contains small scales with (dimensional) amplitudes $O(Fr^{1/2}U^*)$. 
\rev{The scaling law for $H^*$ has} been validated in numerical and laboratory experiments \citep[e.g.][]{HolfordLinden1999,brethouwer2007,Oglethorpeal2013}, with at least suggestive observational evidence of this `layered anisotropic stratified turbulence' (LAST) regime being obtainable 
from seismic oceanography surveys, \rev{as already noted above} \citep{falder2016}. \rev{The scaling law for $W^*$ was first tentatively identified by \citet{maffioli2016}. More recently \citet{Garaudal2024b} measured the root-mean-squared vertical velocity of the flow in DNS of strongly stratified simulations where the turbulence is spatio-temporally intermittent. Using different weight functions to distinguish averages taken within and outside turbulent regions of the flow, they confirmed that $W^* \propto Fr^{1/2} U^*$ within turbulent patches.
In quiescent regions
of the flow outside these turbulent patches, where small-scale fluctuations are suppressed, \citet{Garaudal2024b} found that $W^* \propto FrU^*$, consistent with the predictions of \citet{billant2001} and \citet{brethouwer2007}.}

The scenario in which $Fr \ll 1$, $Re_b \ge O(1)$ and $\alpha \ll Pe_b \ll 1$ is an intermediate regime where the mean flow is diffusive while the fluctuations are not. 
In this regime, we also find that $\alpha = Fr$. The slow-fast system of equations \eqref{eqn:multiscale_cont}, \eqref{eqn:multiscale_u},  \eqref{eqn:intermediate_b}, and \eqref{eqn:intermediate_w}, expressed in the original variables \rev{$(x,y,z,t)$}, becomes the following quasilinear system. \\
\noindent \textit{Mean flow equations:}
\begin{subequations}\label{eqn:intermediate_multiscale}
\begin{alignat}{2}
 \pd{\boldsymbol{\overline{u}}_{\perp} }{t} +  \boldsymbol{\overline{u}}_{\perp} \cdot \nabla\rev{_\perp} \boldsymbol{\overline{u}}_{\perp} \rev{+ \bar{w} \pd{\boldsymbol{\bar{u}}_\perp}{z} } &= - \nabla_{\perp} \overline{p} - \pd{}{z} \left( \overline{w' {\boldsymbol u'}_\perp} \right) +  \frac{1}{Re} \pd{^2 \boldsymbol{\overline{u}}_{\perp}}{z^2} + \rev{\bar{\boldsymbol{F}}_h}, \\
    \pd{\overline{p}}{z} = 0, \qquad & 
    \nabla \cdot \boldsymbol{\overline{u}} = 0 , \tag{\ref{eqn:intermediate_multiscale}\textit{b},\textit{c}} \\
      \overline{w} &  = -  \pd{}{z} \left( \overline{w' b'} \right) +  \frac{1}{Pe} \pd{^2\overline{b}}{z^2}.
 \tag{\ref{eqn:intermediate_multiscale}\textit{d}}
\end{alignat}

\noindent \textit{Fluctuation equations:}
\begingroup
\allowdisplaybreaks
\begin{alignat}{2}
     \pd{\boldsymbol{u}_{\perp}'}{t} +  \boldsymbol{\overline{u}}_{\perp} \cdot \nabla_{\perp} \boldsymbol{u}_{\perp}' + w' \pd{\boldsymbol{\overline{u}}_{\perp}}{z} &= -  \nabla_{\perp} p' + \frac{1}{Re} \nabla^2 \boldsymbol{u}_{\perp}', \\
     \pd{w'}{t} + \boldsymbol{\overline{u}}_{\perp} \cdot \nabla_{\perp} w' &= - \pd{p'}{z} + \frac{b'}{Fr^2} + \frac{1}{Re}  \nabla^2  w', \\ 
	\nabla \cdot \boldsymbol{u}'&= 0, \\
         \pd{b'}{t} +  \boldsymbol{\overline{u}}_{\perp} \cdot \nabla_{\perp} b'  + w' &= \frac{1}{Pe} \nabla^2 b'   .   
 \end{alignat}
 \endgroup
 \end{subequations}

These equations are valid for $Re \ge Fr^{-2}$ (so $Re_b \ge O(1)$) and 
$Fr^{-1} \ll Pe \ll Fr^{-2}$ (so $\alpha \ll Pe_b \ll 1$). By writing them in the original isotropic variables, we now see that the correct quasilinear equations at this order are almost the same as in the \citet{chini2022} regime, except that 
terms in $\bar{b}$ are dropped (in the mean hydrostatic balance and in the buoyancy perturbation equation) because they are formally of higher order. In addition, the mean buoyancy equation takes the LPN form of \rev{\citet{lignieres1999}}, modified by the fluctuation-induced buoyancy flux. 
\revtwo{Given that the mean fields are independent of fast variables, the fluctuation subsystem again can be treated as a (2D) eigenvalue problem on fast scales that, in this case, admits as solutions linear instability modes of low-$Pr$ parallel shear flows \citep[e.g.,][]{Jones1977,lignieres1999shear}.}

Dimensionally, the vertical length scale and vertical velocity scale are the same as in the non-diffusive regime. The buoyancy field, however, is now dominated by large scales if $Pe_b \ge \alpha^{1/2}$ and by small scales if $Pe_b \le \alpha^{1/2}$. Testing these scaling laws numerically will be very difficult, unfortunately, because the range of the intermediate region is very small (holding $Pe$ constant while varying $Fr^{-1}$ or vice-versa) unless $Pr$ is itself very small. 

Finally, the case where $Fr\ll 1$, $Re_b \ge O(1)$ and $Pe_b \ll \alpha$ corresponds to a fully diffusive regime in which both the mean flow and fluctuations are dominated by diffusion and satisfy the LPN balance derived by \citet{lignieres1999}. In this regime, we have demonstrated that $\alpha = (Fr^2/Pe)^{1/3}$. The slow-fast system of equations \eqref{eqn:multiscale_cont}, \eqref{eqn:multiscale_u}, \eqref{eqn:lowPeb_b} and \eqref{eqn:lowPeb_w}, written in the original variables \rev{$(x,y,z,t)$,} is given below. \\
\noindent \textit{Mean flow equations:} 
\begin{subequations}\label{eqn:lowPe_multiscale}
\begin{alignat}{2}
 \pd{\boldsymbol{\overline{u}}_{\perp} }{t} +  \boldsymbol{\overline{u}}_{\perp} \cdot \nabla\rev{_\perp} \boldsymbol{\overline{u}}_{\perp} \rev{+ \bar{w} \pd{\boldsymbol{\bar{u}}_\perp}{z} } &= - \nabla_{\perp} \overline{p} - \pd{}{z} \left( \overline{w' {\boldsymbol u'}_\perp} \right) +  \frac{1}{Re} \pd{^2 \boldsymbol{\overline{u}}_{\perp}}{z^2} +
\rev{\bar{\boldsymbol{F}}_h}, \\
    \pd{\overline{p}}{z} = 0, \qquad 
    \nabla \cdot \boldsymbol{\overline{u}} &= 0 , \tag{\ref{eqn:lowPe_multiscale}\textit{b},\textit{c}} \\
      \overline{w} &  =  \frac{1}{Pe} \pd{^2\overline{b}}{z^2}. \tag{\ref{eqn:lowPe_multiscale}\textit{d}} 
\end{alignat}

\noindent \textit{Fluctuation equations:}
\begingroup
\allowdisplaybreaks
 \begin{alignat}{2}
     \pd{\boldsymbol{u}_{\perp}'}{t} +  \boldsymbol{\overline{u}}_{\perp} \cdot \nabla_{\perp} \boldsymbol{u}_{\perp}' + w' \pd{\boldsymbol{\overline{u}}_{\perp}}{z} &= -  \nabla_{\perp} p' + \frac{1}{Re} \nabla^2 \boldsymbol{u}_{\perp}', \\
     \pd{w'}{t} + \boldsymbol{\overline{u}}_{\perp} \cdot \nabla_{\perp} w' &= - \pd{p'}{z} + \frac{b'}{Fr^2} + \frac{1}{Re}  \nabla^2  w', \\ 
	\nabla \cdot \boldsymbol{u}'&= 0 ,\\
          w' &= \frac{1}{Pe} \nabla^2 b' .     
 \end{alignat}
 \endgroup
 \end{subequations}

This system is valid for $Re_b \ge O(1)$, and $Pe_b \ll \alpha \ll 1$. As $\alpha = (Fr^2/Pe)^{1/3}$, these constraints are equivalent to $Fr^2 \ll Pe$ (so that $\alpha \ll 1$), $Pe \ll Fr^{-1}$ (so that $Pe_b \ll \alpha$), and $Pe \ge Pr^3 Fr^{-4}$ (so that $Re_b \ge O(1)$). 

Dimensionally, 
the characteristic vertical length scale $H^* = \alpha L^* =  (Fr^2/Pe)^{1/3} L^* = (U^* \kappa^*/N^{*2})^{1/3}$. The characteristic vertical velocity $W^* = \alpha^{1/2} U^* =(Fr^2/Pe)^{1/6} U^* = (U^{*7}\kappa^*/N^{*2} L^{*3})^{1/6}$ and is dominated by small-scale fluctuations. The characteristic buoyancy scale is $Pe (Fr^2/Pe)^{5/6} L^* N^{*2} = (U^{*11/6} \kappa^{*-1/6} N^{*-5/3} L^{*-3/2}) L^* N^{*2} $
and is dominated by small-scale fluctuations as well. These scalings have been validated by the DNS of \citet{cope2020} \rev{in fully turbulent flows. \citet{Garaudal2024b} showed that they also remain valid within turbulent patches for DNS that are in the intermittent regime of \citet{cope2020}. In quiescent regions
outside of the turbulent patches, predictions from the single-scale equations apply \citep[c.f. \S~\ref{sec:lowPrani} and][]{Skoutnev2023}.}

Crucially, we find that in all three regimes the characteristic vertical velocity $W^* = \alpha^{1/2} U^*$ is significantly larger than that predicted from the \rev{single-scale} equations given in \S~\ref{sec:highPrani} and  \S~\ref{sec:lowPrani}, where $W^* = \alpha U^*$. This larger scaling has implications for turbulent vertical transport of buoyancy and passive scalars in stellar interiors (see \S~\ref{sec:conclusions}).

Finally, note that in all of these regimes, we have assumed $Re_b \ge O(1)$, which then implies that $Re_b \gg \alpha$ since $\alpha \ll 1$. Recalling that the fluctuation equations have an effective Reynolds number $Re_b/\alpha$, this condition is necessary to ensure that the small-scale fluctuations can develop without being suppressed by viscosity, and is therefore key to the multiscale expansions derived here. When $Re_b < 1$, viscous effects become important and could strongly affect our conclusions. We do not pursue this issue further here, deferring discussion of viscous regimes to \rev{future work.} 
\rev{However, we note that \citet{Garaudal2024b} confirmed that $Re_b = O(1)$ correctly predicts the boundary of the region of parameter space where the multiscale equations and their predicted scalings apply.}


\rev{While this study focuses on the scaling relationships that emerge from the derivation of these multiscale equations and their implications for turbulence regimes (see \S~\ref{sec:turbulent_regimes} below), solving these equations is\rev{, of course,} also an area of interest not just from a scientific but also an algorithmic standpoint. Accordingly, using} existing algorithms for slow-fast systems \citep{michel2019,chini2022,ferraro2022}, our ongoing work involves numerically solving \rev{the multiscale PDEs}
for a range of values of $Re_b$, which \rev{will} allow numerical identification of \rev{the parameter} values at which these behaviours \rev{are realised} \citep{shah2022}. 

\begin{figure}
    \centering
    \includegraphics[width=0.6\textwidth]{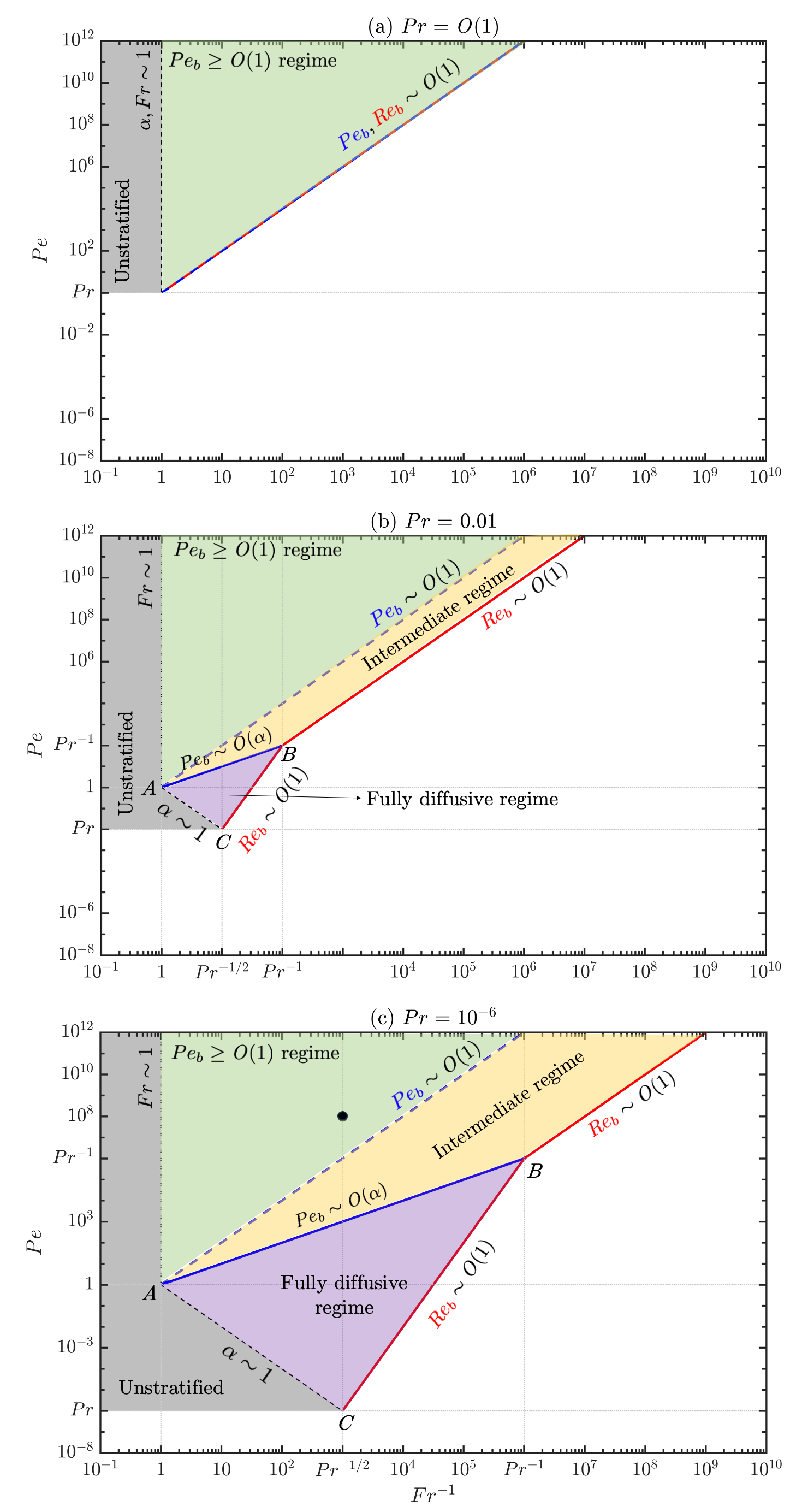}
    
    \caption{Regime diagram for a fluid with \rev{(a)} $Pr = 1$\rev{, (b) $Pr = 10^{-2}$} and \rev{(c)} $Pr=10^{-6}$. The Péclet number is on the vertical axis and inverse Froude number is on the horizontal axis (such that stratification increases to the right). 
    \rev{The unstratified regime is} shown in grey. Regions in white are viscously controlled. 
    \rev{The $Pe_b \ge O(1)$ regime} is marked in green. 
    In each panel, the blue dashed line marks the transition $Pe_b = O(1)$ where the mean flow becomes diffusive, and the red solid lines mark the viscous transition where $Re_b = O(1)$. 
    With $Pr \ll 1$ (panel\rev{s (b) and (c)}), 
    the solid blue line marks 
    $Pe_b = O(\alpha)$ where the small-scale fluctuations become diffusive. 
    A region of parameter space opens up between $Pe_b =O(1)$ and $Re_b =O(1)$, where the two new regimes identified in this work exist: the intermediate regime 
    \rev{(}yellow\rev{)}, and the fully diffusive regime 
    \rev{(}purple\rev{)}. 
    \rev{(c)} A parameter set indicative of the solar tachocline is marked with a black circle, using typical parameters given by \citet{garaud2020}  ($Pe=10^8$, $Fr^{-1}=10^{3}$). }
    \label{fig:regime_diagram}
\end{figure}

\subsection{Regimes of stratified stellar turbulence}
\label{sec:turbulent_regimes}

Our findings partition parameter space 
into various regimes of stratified turbulence, which are illustrated in figure~\ref{fig:regime_diagram} for \rev{(a)} a $Pr=1$ fluid \rev{, (b) a $Pr=10^{-2}$ fluid} and \rev{(c)} a $Pr=10^{-6}$ fluid. \rev{Panel (b)} is typical \rev{of liquid metals and the bottom} of some stellar interiors \citep[see][]{garaud2021}.
In \rev{all} panels, the horizontal axis shows the inverse Froude number $Fr^{-1}$, so that  stratification increases to the right. The vertical axis shows the outer scale P\'eclet number $Pe$, so that diffusive effects decrease upward. 
In both panels, the grey region shows where \rev{$\alpha = O(1)$}, 
so the large-scale flow is isotropic \rev{and effectively unstratified}; this regime is not the focus of our study.
For stronger stratification, the large-scale flow becomes anisotropic, and the possible regimes of stratified turbulence depend on $Pr$.

For $Pr = O(1)$, shown in the upper plot, the partitioning of parameter space for $Fr^{-1} \gg 1$ is straightforward. When $Pe_b, Re_b \ge O(1)$ (green region), viscosity and diffusion play a secondary role in both the mean and fluctuation dynamics. \rev{\citet{billant2001,lindborg2006} put forward the basis for t}his regime\rev{, for which asymptotic theory was developed} by \citet{chini2022}. By contrast, if $Pe_b, Re_b \ll 1$ (white region) then both effects \rev{strongly} influence the flow dynamics. 
The transition takes place when $Re,Pe \approx Fr^{-2}$.

At low $Pr$, diffusion becomes important long before viscosity does, so the $Pe_b = O(1)$ transition is distinct from the $Re_b = O(1)$ transition. 
This distinction opens up parameter space to the two new regimes discussed in this work: the intermediate regime (yellow region, the dynamics of which are described in  \S~\ref{sec:intermediate}) and the fully-diffusive regime (purple region, the dynamics of which are described in \S~\ref{sec:meanLowPeb_fluctLowPe_b}). 
We now clearly see that the fully-diffusive regime is confined to a triangle in log-log parameter space for a given Prandtl number. 
It is delimited from above by the intermediate regime (yellow region), from below by the isotropic 
regime (grey region), and from the right by the viscous regime (white region). 
More specifically, it is bounded by the points A $ \, (Fr^{-1} = 1, Pe = 1)$, B $ \, (Fr^{-1} = Pr^{-1}, Pe = Pr^{-1})$ and C $ \, (Fr^{-1} = Pr^{-1/2}, Pe = Pr)$ and thus becomes increasingly wide as $Pr$ decreases, but shrinks towards the point A as $Pr \rightarrow 1$ \rev{(e.g., compare panels (b) and (c))}. \rev{Note that the isotropic (grey) region in panels (b) and (c) contains an additional triangular area which arises from the scaling relationship \eqref{eq:lowPebalpha}: $\alpha = 1$ corresponds to $Pe = Fr^{2}$ there, in contrast to the $Pr \simeq \mathit{O}(1)$ case in panel (a) where $\alpha = 1$ implies $Fr = 1$. In all panels, the non-turbulent regime where $Re < 1$ is marked by $Pe < Pr$, which has been left white.}


A major implication of our results is that, depending on the choice of $Pe$, different regimes are encountered as the stratification increases.
We now discuss horizontal transects through \rev{the bottom panel of}  figure~\ref{fig:regime_diagram}, for different values of $Pe$.
We first consider the case   $Pr < Pe  < 1$, which is the regime discussed in \citet{cope2020}. As $Fr^{-1}$ increases \rev{(while holding $Pe$ and $Re$ constant)}, our model predicts that the turbulence ought to be isotropic until $Fr^{-1} = Pe^{-1/2}$ (interestingly, because diffusion partially relaxes the effects of stratification when $Pe \ll 1$). As $Fr^{-1}$ continues to increase, the turbulence enters the fully diffusive anisotropic regime, and remains in that regime until $Fr^{-1} = (Pe/Pr^3)^{1/4}$, at which point viscosity begins to affect the mean flow. This series of regime transitions is qualitatively consistent with \rev{that} observed in the low Prandtl number DNS of \citet{cope2020}. 

At the other extreme,  let us consider a transect for $Pe > Pr^{-1}$, which is the case considered by \citet{garaud2020}, who primarily analysed simulations for which $Pe = 60$ and $Pr = 0.1$. This regime is relevant for strongly sheared layers in stellar interiors, such as the solar tachocline. For moderate stratification, namely $1 \le Fr^{-1} \le Pe$, our analysis shows that the turbulence is expected to be both anisotropic and non-diffusive, and its properties should be captured by the model of \citet{chini2022}. As stratification increases past $Fr^{-1} = Pe$, the turbulence is predicted to enter the intermediate regime, where the mean flow is dominated by diffusion but the fluctuations are not. Beyond $Fr^{-1} = \sqrt{Pe /Pr}$, viscous effects should become important.
Note how, at these large values of the Péclet number, the fully diffusive inviscid regime discussed in \S~\ref{sec:meanLowPeb_fluctLowPe_b} is not accessible. Instead, viscosity begins to influence the mean flow before diffusion influences the fluctuations. 

This series of regime transitions is qualitatively consistent with the simulations reported in \citet{garaud2020}.
Specifically, she found that the turbulence is mostly isotropic for $Fr^{-1} < 1$, then becomes anisotropic with little effect of diffusion for intermediate values of $Fr^{-1}$. 
However, her empirically-derived scaling laws 
($H^* \propto Fr^{2/3} L^*$, $W^* \propto Fr^{2/3} U^*$)
do not match those predicted by the \citet{chini2022} theory. 
\rev{\citet{Garaudal2024b} have now resolved this discrepancy by showing that the apparent $W^* \propto Fr^{2/3} U^*$ scaling law is an artifact of the coexistence of turbulent patches where $W^* \sim Fr^{1/2} U^*$ and laminar regions where $W^* \sim Fr U^*$, whose volume fraction respectively decreases and increases as stratification increases.}

For even larger values of $Fr^{-1}$, \citet{garaud2020} found that diffusion becomes important, and that the dynamics are governed by the LPN equations. 
However, at this point the turbulence is also 
\rev{viscously affected,} a situation qualitatively consistent with \rev{the right-most regime transition in}
figure~\ref{fig:regime_diagram}. 

\rev{T}he case where $1 < Pe < Pr^{-1}$ is relevant for weaker shear layers in stellar interiors. As $Fr^{-1}$ increases, our multiscale analysis predicts that the turbulence ought to be isotropic until $Fr^{-1} = 1$, at which point the regime analysed by \citet{chini2022} sets in and both the mean and fluctuation dynamics are non-diffusive. As the stratification continues to increase, the vertical eddy scale decreases gradually until the mean flow becomes dominated by diffusion at $Fr^{-1} = 
\sqrt{Pe}$ and the intermediate regime is manifest. The flow dynamics remain essentially unchanged in that regime until diffusion also begins to affect the fluctuations as well, at $Fr^{-1} = 
Pe$, at which point the turbulence becomes fully diffusive. Finally, viscous effects begin to be important when $Fr^{-1} = (Pe/Pr^3)^{1/4}$.  
To date, no DNS have been published probing this range of $Pe$. Verifying the existence of these successive transitions with DNS is one of our areas of active work.

\subsection{Mathematical considerations}\label{sec:mathsy_discussion} 

Having constructed a map of parameter space based on the results of our multiscale analysis, we now discuss important consequences and caveats of the resulting reduced models. As in \citet{chini2022}, we find that the reduced equations in each regime form a closed set describing the concurrent evolution of a highly anisotropic large-scale mean flow together with isotropic small-scale fluctuations. Crucially, the fluctuation equations derived in all regimes identified 
 are linear in the fluctuation fields, but feed back nonlinearly on the mean flow evolution. The reduced models thus all have a quasilinear form, which emerges self-consistently from the asymptotic analysis rather than being imposed {\it a priori}.
This emergent quasilinearity in the slow-fast limit therefore appears to be an inherent property of stratified turbulence regardless of the Prandtl number. Our findings can be used as a theoretical basis not only for using the quasilinear approximation to create reduced models of stratified turbulence, but also to know precisely which terms to keep in each region of parameter space \citep[see, e.g,][]{MarstonTobias2023}. Of course, the fundamental premise upon which the analysis is predicated is that the stratification drives scale-separated dynamics characterized by the anisotropic large-scale and roughly isotropic small-scale flow structures. To the extent that spectrally non-local interactions between these disparate scales of motion play a crucial role in stratified turbulence, the conclusions and predictions of our multiscale analysis are likely to be valid.

In quasilinear systems subject to fast instabilities, nonlinear saturation requires the feedback from the finite-amplitude fluctuations to render the mean fields marginally stable to disturbances of any horizontal wavenumber.
It is straightforward to verify that this condition of approximate marginal stability indeed characterizes each regime. This \rev{deduction} is consistent with the empirical observation of `self-organized criticality' noted in \S~\ref{sec:intro}, giving further credence to the appropriateness of the multi-scale approach used here.
In the $Pe_b \ge O(1)$ and intermediate regimes, the fluctuation equations are non-diffusive at leading order, and therefore describe the growth (or decay) of perturbations due to a standard vertical shear instability. The dimensional vertical shear $S^*$ of the mean flow \rev{(which emerges once velocity layers develop)} can be estimated to be roughly $U^*/H^*$ since $|\boldsymbol{\bar  u}| = O(1)$. 
The 
gradient Richardson number 
\begin{equation}
J = \frac{N^{*2}}{S^{*2}} = O\left(\frac{N^{*2}H^{*2}}{U^{*2}}\right) = O\left(\frac{\alpha^2}{Fr^2}\right)  = O(1) 
\end{equation}
in these regimes, which is \rev{indicative of marginal stability according} to 
the Richardson and Miles--Howard criteria \citep{Richardson1920,Miles1961,Howard1961}. Note that \citet{Garaudal2024a} recently verified that $J$ is indeed $O(1)$ in DNS of stratified turbulence at $Pe_b = O(1)$. 

In the $Pe_b \ll \alpha$ regime, by contrast, the fluctuation equations are inherently diffusive, and one therefore expects the vertical shear instability to be of the diffusive kind \citep{zahn1974,Jones1977,Lignieresal1999}. It has been shown, at least for sinusoidal shear flows \citep{GaraudGallet2015}, that the condition for marginal linear stability is $J Pe_S = O(1)$ where $Pe_S = U^* H^*/ \kappa^*$ is the P\'eclet number based on the vertical shear profile. We can easily check that the scalings found in \S~\ref{sec:meanLowPeb_fluctLowPe_b} satisfy this criterion. Indeed, 
\begin{equation}
   J Pe_S  =  O\left(\frac{N^{*2}H^{*2}}{U^{*2}}\frac{U^* H^*}{\kappa^*} \right) =  O\left(\alpha^3 \frac{Pe}{Fr^2} \right)  = O(1) ,
\end{equation}
as required. Taken together with the empirical scaling laws obtained from the DNS data of \citet{cope2020}, this evidence confirms that \eqref{eq:lowPebalpha} is the correct expression for $\alpha$ in stratified turbulence at very low $Pe_b$. 

A somewhat less intuitive result of our analysis is the requirement that $\partial \bar p_{00}/\partial \zeta = 0$, while $\bar p_{00} = O(1)$, in both \rev{the} intermediate and fully diffusive regimes. Physically, this condition can be understood by noting that in the $Pe_b \ll 1$ scenario, departures from the mean background stratification formally are exceedingly small (owing to the strong thermal diffusion), which explains why they do not affect the assumed {\it background} hydrostatic balance. Furthermore, recalling that $\zeta$ has been rescaled by the vertical length scale $H^* = \alpha L^*$ with $\alpha \ll 1$, we note that $\partial_\zeta \overline{p}_{00} = 0$ only applies on that scale, and does not preclude $\bar p_{00}$ from potentially varying on larger scales. 
Indeed, the possibility exists that \emph{two} vertical scales may be incorporated into the dynamics, with the leading-order mean pressure being hydrostatically coupled to the leading-order mean buoyancy on the larger vertical scale, although we leave exploration of that possibility for future work.
\rev{Alternatively, as discussed in \S~3.3, higher-order mean pressure terms, which couple to the mean buoyancy, can be incorporated using a composite mean vertical momentum equation. Then, in the singular limit of no fluctuations, the sets of multiscale equations for the intermediate \eqref{eqn:intermediate_multiscale} and fully diffusive \eqref{eqn:lowPe_multiscale} regimes recover the single-scale equations in \S~2.2.} 
Regardless, we emphasize that buoyancy anomalies \emph{do} affect the fluctuation dynamics, which in turn modifies the mean flow.


\section{Conclusions}\label{sec:conclusions}

In this study, inspired by numerical evidence of scale separation and flow anisotropy, we have conducted a formal multiscale asymptotic analysis of the Boussinesq equations governing the dynamics of strongly stratified turbulence at low Prandtl number.
A key outcome of our work is a new map of parameter space, shown in figure~\ref{fig:regime_diagram}, that demarcates different regimes of stratified turbulence. 
Crucially, we find that new regions of parameter space open up at low Prandtl number in which diffusive turbulent flows exist, scenarios that are not possible at $Pr = \mathit{O}(1)$. 
For each of these new regimes, scaling laws for the vertical velocity and vertical length scale of turbulent eddies naturally emerge from the analysis.
These scaling laws are summarized in \S~\ref{sec:turbulent_regimes} and recover previous findings by \citet{chini2022} and \citet{cope2020} in appropriate distinguished limits. Finally, recent work by \citet{Garaudal2024a} has demonstrated numerically that the presence of a mean vertical shear (ignored in this work) has little impact on these scalings as long as its amplitude is smaller than the small-scale emergent vertical shear $U^*/H^*$. This finding can be proved more formally using the asymptotic tools developed here, suggesting that the new regime diagram is robust (at least if other effects such as rotation and magnetic fields are absent; see below).

These results have important implications for low $Pr$ fluids, such as liquid metals (where $Pr \sim 0.01-0.1$), planetary interiors (where $Pr \sim 0.001-0.1$) and stellar interiors (where $Pr \sim 10^{-9}-10^{-2}$).
In particular, the \rev{derived} scaling laws enable us to propose simple parameterizations for the turbulent diffusion coefficient ($D^*$) \rev{of a passive scalar} in each identified regime by multiplying the characteristic vertical length and vertical velocity scales. 
\begin{itemize}
    \item In the $Pe_b \geq \mathit{O}(1)$ and the intermediate regimes, 
    \begin{equation}
        D^* \propto H^* W^* \propto Fr^{3/2} L^* U^* \propto U^{* \, 5/2}/(L^{* \, 1/2}N^{*3/2}).
    \end{equation}
    \item In the fully diffusive regime,  
    \begin{equation}\label{eqn:turb_diff}
        D^* \propto H^* W^* \propto (Fr^2/Pe)^{1/2} L^* U^* \propto U^{* \, 3/2} \kappa^{*1/2} / (L^{* \, 1/2} N^*).
    \end{equation}
\end{itemize}

As such, our work challenges current understanding of stratified turbulence in stars.
Indeed, the most commonly used model for stratified turbulence in stellar evolution calculations is the model of \citet{zahn1992}, which proposes a vertical turbulent diffusivity equivalent to \eqref{eqn:turb_diff}.
However, we have demonstrated that this expression is only valid in a relatively small region of the parameter space (see figure~\ref{fig:regime_diagram}). 
In particular, Zahn's model assumes that the turbulence is always fully diffusive, but this is not the case in the intermediate and $Pe_b \geq \mathit{O}(1)$ regimes, which are appropriate for more strongly sheared fluid layers such as the solar tachocline. 
Zahn's model also assumes that viscosity is negligible, which is not the case for sufficiently stratified flows.

\rev{Our multiscale analysis can also be used to estimate the magnitude of the  dimensional turbulent buoyancy flux ${\cal B}^* = |\overline{w'b'}| N^{*2} U^* L^*$, using the characteristic sizes of $w'$ and $b'$ found in each regime. 
\begin{itemize}
    \item In the $Pe_b \ge O(1)$, $Re_b \gg 1$ regime, $w' = O(\alpha^{1/2})$ and $b' = O(\alpha^{3/2})$ with $\alpha = Fr$, so
    \begin{equation}
        {\cal B}^* \propto \alpha^{2}  N^{*2} U^* L^* \propto U^{*3}/L^*.
\end{equation}
\item In the $Pe_b \ll \alpha, Re_b \gg 1$ regime, $w' = O(\alpha^{1/2})$ and $b' = O(Pe_b \, \alpha^{1/2})$ with $\alpha = (Fr^2/Pe)^{1/3}$ so we also find in this case that 
\begin{equation}
    {\cal B}^* \propto Pe_b \, \alpha \,  N^{*2} U^* L^* \propto Pe \, \alpha^3  N^{*2} U^* L^* \propto U^{*3}/L^*.
\end{equation}
\end{itemize}
This demonstrates that both regimes have a flux coefficient $\Gamma = {\cal B}^* / \varepsilon^*$ (where $\varepsilon^* \propto U^{*3}/L^*$ estimates the kinetic energy dissipation rate) and a mixing efficiency $\eta = {\cal B}^*/({\cal B}^* + \varepsilon^*)$ that are $\mathit{O}(1)$ and  independent of stratification \rev{(provided the stratification is sufficiently strong that $\alpha\ll 1$).} This \rev{conclusion} is consistent with \rev{the analysis of} \citet{maffioli2016b}
and with the DNS of \citet{cope2020} at  $Pr \ll 1$. }


\rev{Further} work \rev{is} needed to investigate the effects of rotation and magnetization, which are also important in stellar and planetary interiors. Both effects are likely to stabilize the horizontal turbulence to some extent, which will therefore affect the emergent vertical shear instability as well. 
Nonetheless, this work has already demonstrated the power of formal multiscale asymptotic analysis for discovering and validating the existence of new regimes of stratified turbulence in stellar and planetary interiors.

\section*{Acknowledgements}

\noindent The authors gratefully acknowledge the Geophysical Fluid Dynamics Summer School (NSF 1829864), particularly the 2022 program.
P.G. acknowledges funding from NSF AST-1814327.
K.S. acknowledges funding from the James S. McDonnell Foundation.
For the purpose of open access, the authors have applied a Creative Commons Attribution (CC BY) licence to any Author Accepted Manuscript version arising from this submission.

\section*{Declaration of Interests}

\noindent The authors report no conflict of interest.

\bibliographystyle{apalike}
\bibliography{Shahetal_arXiv_submission_RegimesStratifiedTurbulenceLowPr_Revision2}

\end{document}